\documentclass[twocolumn,superscriptaddress,amsmath,amssymb,aps,pra, sort&compress]{revtex4-1}
\usepackage{graphicx}
\usepackage{dcolumn}
\usepackage{bm}

\usepackage{color}

\usepackage{graphicx}
\usepackage{comment}
\usepackage{times,color}
\usepackage[colorlinks,urlcolor=blue,citecolor=blue,linkcolor=blue]{hyperref}
\usepackage{physics}
\usepackage{mathtools}
\usepackage{mathrsfs}
\usepackage{booktabs}

\newcommand{\kb}{k_{\rm{B}}}

\newcommand{\eb}{\varepsilon_b}

\newcommand{\down}{\downarrow}
\newcommand{\up}{\uparrow}
\renewcommand{\k}{{\bf k}}
\newcommand{\p}{{\bf p}}
\newcommand{\q}{{\bf q}}
\newcommand{\0}{{\bf 0}}

\newcommand{\ef}{\varepsilon_F}

\newcommand{\kf}{k_F}

\newcommand{\ep}{\epsilon_{\p}}

\newcommand{\ek}{\epsilon_{\k}}
\newcommand{\ekmed}{\epsilon_{\k}^{\rm med}}
\newcommand{\ekpqmed}{\epsilon_{\k+\q}^{\rm med}}

\newcommand{\ekpmq}{\epsilon_{\p-\q}}

\newcommand{\beq}{\begin{equation}}
\newcommand{\eeq}{\end{equation}}

\usepackage[normalem]{ulem}   

\begin{document}

\title{Repulsive Fermi and Bose Polarons in Quantum Gases}



\author{Francesco Scazza}
\email[] {francesco.scazza@units.it}
\affiliation{Department of Physics, University of Trieste, 34127 Trieste, Italy}
\affiliation{Istituto Nazionale di Ottica del Consiglio Nazionale delle Ricerche (CNR-INO), 50019 Sesto Fiorentino, Italy}
\affiliation{\mbox{European Laboratory for Nonlinear Spectroscopy (LENS), University of Florence, 50019 Sesto Fiorentino, Italy}}

\author{Matteo Zaccanti}
\email[] {zaccanti@lens.unifi.it}
\affiliation{Istituto Nazionale di Ottica del Consiglio Nazionale delle Ricerche (CNR-INO), 50019 Sesto Fiorentino, Italy}
\affiliation{\mbox{European Laboratory for Nonlinear Spectroscopy (LENS), University of Florence, 50019 Sesto Fiorentino, Italy}}

\author{Pietro Massignan}
\email[] {pietro.massignan@upc.edu}
\affiliation{Departament de F\'isica, Universitat Polit\`ecnica de Catalunya, Campus Nord B4-B5, 08034 Barcelona, Spain}

\author{ Meera M. Parish}
\email[] {meera.parish@monash.edu}
\affiliation{School of Physics and Astronomy, Monash University, Victoria 3800, Australia}
\affiliation{ARC Centre of Excellence in Future Low-Energy Electronics Technologies, Monash University, Victoria 3800, Australia}

\author{ Jesper Levinsen}
\email[] {jesper.levinsen@monash.edu}
\affiliation{School of Physics and Astronomy, Monash University, Victoria 3800, Australia}
\affiliation{ARC Centre of Excellence in Future Low-Energy Electronics Technologies, Monash University, Victoria 3800, Australia}

\begin{abstract}
Polaron quasiparticles are formed when a mobile impurity is coupled to the elementary excitations of a many-particle background. In the field of ultracold atoms, the study of the associated impurity problem has attracted a growing interest over the last fifteen years. Polaron quasiparticle properties are essential to our understanding of a variety of paradigmatic quantum many-body systems realized in ultracold atomic gases and in the solid state, from imbalanced Bose-Fermi and Fermi-Fermi mixtures to fermionic Hubbard models. In this topical review, we focus on the so-called repulsive polaron branch, which emerges as an excited many-body state in systems with underlying attractive interactions such as ultracold atomic mixtures, and is characterized by an effective repulsion between the impurity and the surrounding medium. We give a brief account of the current theoretical and experimental understanding of repulsive polaron properties, for impurities embedded in both fermionic and bosonic media, and we highlight open issues deserving future investigations.
\end{abstract}


\maketitle

\section{Introduction}

Understanding the fate of an impurity particle immersed within a complex medium represents a paradigmatic problem in quantum physics \cite{Mahan2000book}, encompassing a variety of physical scenarios and spanning an enormous range of energies: from ultracold atomic gases \cite{Chevy2010,Massignan_Zaccanti_Bruun,Levinsen2Dreview}, helium liquids \cite{Bardeen1967,Lemeshko2018} and solid-state materials \cite{LandauPekar,Froehlich1954,Feynman1955,Alexandrov2010}, 
all the way up to nuclear and quark matter \cite{Klimov1981,Weldon1989,Nakano2020}.
The interest in such $(N+1)$ many-body systems is two-fold: On the one hand, a primary goal is to characterize how the interactions with the surrounding bath turn the impurity particle into a \textit{quasiparticle} with modified static and dynamical properties, such as a renormalized energy, a finite lifetime and an effective mass. 
On the other hand, understanding how the bare particle is effectively \textit{dressed} by its environment provides important information about the nature of the medium itself, since the impurity can act as a microscopic, local probe for both the excitations and the collective properties of the surrounding material. 

Importantly, a common set of ideas and technical tools can be applied to investigate the impurity problem in seemingly disparate setups, the investigation of one system yielding information on another. In particular, the so-called Fermi polaron -- a mobile quantum impurity embedded in a degenerate Fermi gas -- has been  successfully investigated and characterized both in ultracold atomic mixtures~\cite{Schirotzek2009,Nascimbene2009,Kohstall2012,Koschorreck2012,Zhang2012,Wenz2013,Ong2015,Cetina2015,Cetina2016,Scazza2017,Mukherjee2017,Yan2019a,Oppong2019,Ness2020,Adlong2020,Fritsche2021} and in atomically thin semiconductors~\cite{Sidler2017,Wang2018}, using a single theoretical framework and relying on similar experimental methods. This has recently stimulated much cross-fertilization between the two fields (see, e.g., Ref.~\citenum{Bastarrachea2021}). In addition, the original polaron scenario of a single electron interacting with phonons in a crystal~\cite{LandauPekar} has now been extended to quantum impurities immersed in a Bose Einstein condensate -- the so-called Bose polaron~\cite{Catani2012,Hu2016,Jorgensen2016,Yan2019,Skou2021}.

Ultracold atoms represent a particularly appealing playground for the exploration of impurity physics, both in and out of equilibrium, owing to their versatility, their clean and isolated nature, as well as their accessible time and length scales. 
Initiated with the investigation of the highly-polarized limit of resonantly interacting Fermi gases across the BCS-BEC crossover, and the observation of the so-called attractive Fermi polaron \cite{Schirotzek2009,Nascimbene2009}, over the last decade a series of groundbreaking experiments have renewed the interest in the $(N+1)$ problem within the field of quantum gases. 
These have already enabled the characterization of the quasiparticle properties of impurity atoms embedded within both Fermi~\cite{Schirotzek2009,Nascimbene2009,Kohstall2012,Koschorreck2012,Zhang2012,Wenz2013,Ong2015,Cetina2015,Cetina2016,Scazza2017,Mukherjee2017,Yan2019a,Oppong2019,Ness2020,Adlong2020,Fritsche2021} and Bose \cite{Hu2016,Jorgensen2016,Ardila2018,Yan2019,Skou2021} environments, for various impurity-to-medium particle mass ratios, and encompassing not only systems in three dimensions (3D), but also  two- \cite{Koschorreck2012,Zhang2012,Oppong2019} and one-dimensional \cite{Palzer2009,Catani2012,Meinert2017}  environments.
Combining the exquisite control over interatomic interactions enabled by magnetic Feshbach resonances \cite{Chin2010} with advanced spectroscopic tools \cite{Torma2016,Vale2021}, quantum gases allow one to prepare both impurity and medium particles in single, well-defined quantum states, and to probe quasiparticle properties with unparalleled accuracy, down to the single-atom level.

As a non-trivial and quite general result, the detailed comparison between experiment and theory has demonstrated that most quasiparticle properties can be accurately modeled, even in the strong coupling regime, using theoretical methods which are much simpler than those required for a quantitative description of balanced atomic mixtures.
For instance, excellent agreement between theory and experiment has been demonstrated for the ground-state properties of a highly imbalanced Fermi mixture, owing to the almost exact cancellation of a large set of high-order Feynman diagrams \cite{Combescot2008,Houcke2019}, and the agreement even extends to the non-equilibrium evolution of impurities immersed in an ultracold Fermi gas following an interaction quench \cite{Cetina2016,Liu2019}.
Simple theories for strongly interacting many-body systems are rare, and the \mbox{$(N+1)$} problem therefore provides an important testbed for improving our understanding of more complex states of highly correlated matter. 
In particular, the study of the extremely polarized case of a single impurity provides accurate information for systems that feature a sizable concentration of minority particles,
the impurity limit exhibiting some of the critical points of the full phase diagram, whose topology we can thus learn about by investigating highly polarized systems \cite{Parish2007}.

\begin{figure}[t]
\begin{center}
\includegraphics[scale=1.4]{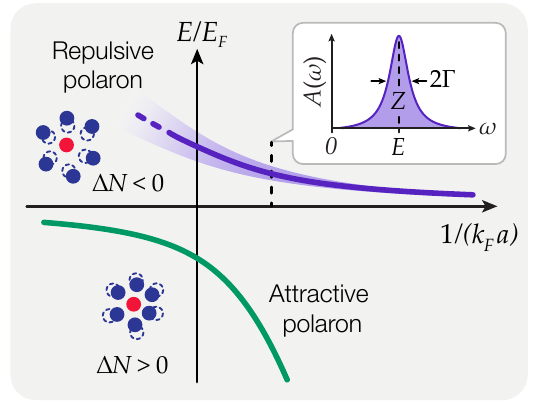}
\caption{Quasiparticle spectrum as a function of interaction strength $1/(k_Fa)$. Attractive (green) and repulsive (purple) polaron energy branches. The shaded area centered around the repulsive polaron energy represents the quasiparticle spectral width $\Gamma$. For widths comparable to its energy, the repulsive polaron ceases to be a well defined \emph{coherent} quasiparticle (dashed line ending). In the case of a fermionic medium, the attractive polaron also stops being well defined at sufficiently large $1/(k_Fa)$, where it undergoes a sharp transition to a dressed molecule quasiparticle~\cite{Prokofev2008}. On the other hand, the ground state of the Bose polaron spectrum does not feature a single-impurity transition. Inset: impurity spectral function $A(\omega)$ at zero momentum for $\omega >0$, i.e.~a vertical cut through the repulsive polaron spectrum at fixed interaction strength. The center of the polaron spectral function denotes the polaron energy $E$, while its half width at half maximum and area relate to the quasiparticle width $\Gamma$ and residue $Z$, respectively.}
\label{fig:schematic}
\end{center}
\end{figure}

In this topical review article, we focus on a specific kind of impurity quasiparticle, termed the \emph{repulsive polaron}, discussing experimental and theoretical progress in the understanding of its highly non-trivial nature from a cold-atom perspective.
Originally introduced to characterize the Stoner instability \cite{Stoner1933} of a gas of itinerant fermions towards a ferromagnetic state \cite{Pilati2010,Chang2011,Massignan_Zaccanti_Bruun}, the repulsive polaron concept is nowadays generically employed to denote any impurity particle dressed by strong repulsive interactions with a surrounding (either fermionic or bosonic) medium. 
While for genuine impurity-medium interparticle repulsion, e.g., the one originating from Coulomb or hard-sphere interaction potentials, the repulsive polaron represents the ground state of the $(N+1)$-system, in the case of van der Waals interactions relevant for cold atomic gases \cite{Chin2010} such repulsive quasiparticles connect to an excited energy branch of the many-body spectrum (see Fig.~\ref{fig:schematic}). 
This is due to the fact that any short-ranged repulsion with a scattering length exceeding the interaction range inherently requires an underlying weakly-bound molecular level into which the system may decay, thereby making the repulsive polaron metastable. 
As a consequence, repulsive quasiparticles in ultracold atomic mixtures represent both a theoretical and experimental challenge, which has stimulated over the last decade an intense debate about the nature of the repulsive branch, with even its existence being questioned \cite{Goulko2016}. On the other hand, this has triggered the rapid development of new theoretical methods and experimental probes, able to trace in real time the quasiparticle formation, decay and decoherence~\cite{Goold2011,Knap2012,Cetina2016,Parish2016,Schmidt2018,Amico2018,Liu2019,Adlong2020,Skou2021}.

Here, we provide a concise overview of the recent advances in this research field. The remainder of the paper is organized as follows: in Section \ref{sec:polarons}, we outline the theoretical basis for the treatment of single-impurity problems in ultracold bosonic and fermionic atomic media; in Section \ref{sec:expt}, we introduce the main experimental probes of polaron quasiparticle properties, especially focusing on the metastable repulsive branch; in Section \ref{sec:qp-lifetime}, we discuss the origin of the repulsive polaron quasiparticle lifetime, reconciling different interpretations for the quasiparticle damping mechanisms found in the literature; finally, in Section \ref{sec:interactions}, we discuss the emergence of long-range impurity-impurity interactions mediated by the medium, thereby linking polaron physics to that of bosonic and fermionic atomic mixtures.

\section{Fermi and Bose polarons} \label{sec:polarons}

We begin by introducing the problem of a single impurity in a quantum medium and providing an overview of the basic theoretical concepts. 
Our focus will be on the case of short-range interactions between the impurity and medium particles, which is appropriate for dilute quantum gases.
For concreteness, we will restrict our attention to three-dimensional (3D) systems, since this is the situation for most of the cold-atom experiments that have been performed thus far. 
However, the phenomenology of the repulsive polaron branch is similar in two dimensions~\cite{Adlong2020}, and thus our discussion is also relevant to exciton polarons in atomically thin semiconductors~\cite{Sidler2017}. 

\subsection{Theoretical description} 
In this review, we will assume that the medium consists of identical particles of mass $m_{\rm med}$ which are in thermal equilibrium at a temperature $T$ and chemical potential $\mu$. The medium can, for instance, correspond to an ideal Fermi gas or a weakly interacting Bose gas. It is governed by a Hamiltonian $\hat H_{\rm med}$ which, in the two cases, is given by either
\begin{align}
\hat H_{\rm Fermi}=\sum_\k (\ekmed-\mu) \hat f^\dag_\k \hat f_\k,
\end{align}
where the creation operator $\hat f_\k^\dag$ satisfies the usual fermionic commutation relations, or
\begin{align}
\label{eq:Bose}
\hat H_{\rm Bose}=\sum_\k (\ekmed-\mu) \hat b^\dag_\k \hat b_\k+\sum_{\k\k'\q} V_B(\q)\hat b^\dag_{\k+\q}\hat b^\dag_{\k'-\q} \hat b_{\k'}\hat b_{\k},
\end{align}
with the bosonic creation operator $\hat b^\dag_\k$. The single-particle dispersion in the medium is $\epsilon_{\k}^{\rm med}=|\k|^2/2m_{\rm med}\equiv k^2/2m_{\rm med}$ at momentum $\k$. To be able to directly compare results for the two cases, we will define a Fermi wave vector $k_F=(6\pi^2 n)^{1/3}$ and Fermi energy $E_F =  k_F^2/2m_{\rm med}$ in terms of the density $n$ of the bath, independent of the medium statistics ($k_F$ and $E_F$ are often labeled $k_n$ and $E_n$ in the Bose polaron literature). In Eq.~\eqref{eq:Bose} for a bosonic medium, we have introduced a boson interaction potential $V_B$ which is assumed to be of short range and characterized by a scattering length $a_B$ which is positive and small, $0<na_B^3\ll1$, to ensure the stability of the Bose gas. 
Here and throughout this review we use units where the volume, Boltzmann's constant $\kb$ and the reduced Planck's constant $\hbar$ are all set to one.

Including the impurity degree of freedom as well as the impurity-medium interactions, we therefore have the total Hamiltonian
\begin{align}\label{eq:Hgen}
\hat H=\hat H_{\rm med}+\sum_{\k}\ek \hat c^\dag_\k \hat c_\k + \sum_{\k,\q} V(\q) \hat \rho_\q \hat c^\dag_{\k+\q}\hat c_\k.
\end{align}
Here, $\hat c^\dag_\k$ is the impurity creation operator and $\ek=k^2/2m$ is the impurity kinetic energy, with $m$ the impurity mass.
The bosonic operator $\hat \rho_\q$ corresponds to medium density operators, and takes the form $\hat \rho_\q=\sum_\k \hat f^\dag_{\k-\q}\hat f_{\k}$ or $\hat \rho_\q=\sum_\k \hat b^\dag_{\k-\q}\hat b_{\k}$ depending on the statistics of the medium. 
Note that we treat the impurity within the canonical ensemble, where we have a fixed number of impurities (one impurity in this case), while we use the grand canonical ensemble for the medium. 

We have written the impurity-medium interactions in the Hamiltonian~\eqref{eq:Hgen} in terms of a generic finite-range potential $V(\q)$ which can in principle describe an actual repulsive interaction such as a soft-sphere potential, as well as the short-range attractive interactions that are the main focus of this review. 
Importantly, regardless of whether $V(\q)$ is attractive or repulsive, the low-energy scattering amplitude between an impurity and a medium atom at relative momentum $\k$ has the \textit{universal} form
\begin{align}
f_s(\k)=-\frac1{1/a+ik},
\label{eq:scatamp}
\end{align}
where $a$ is the $s$-wave scattering length. This allows us to employ  pseudo-potentials for the impurity-medium interactions that simplify calculations and expose the universal physics.  

In the following, 
we take $V(\q)=g$, where the constant $g$ is the coupling strength, and we introduce an ultraviolet cutoff $\Lambda$ --- effectively corresponding to the (inverse) range of the potential --- on the relative collision momenta in all two-body scattering processes. 
We then use the low-energy scattering amplitude in Eq.~\eqref{eq:scatamp} to relate the physical parameter $a$ to the ``bare'' parameters of the model, $g$ and $\Lambda$, yielding the relation
\begin{align} \label{eq:1g1a}
 \frac1g = \frac{m_r}{2\pi a}-\sum_\k^\Lambda\frac1{\ek+\ekmed} =  \frac{m_r}{2\pi a}- \frac{m_r}{2\pi^2} \Lambda,
\end{align}
where $m_r=(1/m+1/m_{\rm med})^{-1}$ is the reduced mass. Here, we immediately see that for repulsive interactions $g > 0$, we have the constraint $\pi/\Lambda > a > 0$, where $\pi/\Lambda$ mimics the range of a repulsive potential. Thus, in this case, the scattering length $a$ is always positive, and it has an upper limit set by the range of the potential itself.

In the case of ultracold atomic gases where the underlying van der Waals interactions are attractive, there are no such restrictions on the scattering length and it can be freely tuned to both positive and negative values by varying an external magnetic field. In particular, in the vicinity of a Feshbach resonance~\cite{Chin2010} it can be made to greatly exceed any other length scale in the problem, attaining the regime of unitarity-limited interactions.
When the scattering length is positive, there exists a shallow bound state between the impurity and a particle from the medium, with binding energy $\eb=1/2m_r a^2$, corresponding to the pole of $f_s(\k)$ at $k=i/a$~\cite{Sakurai1985}.

\begin{table*}[t]
\centering
\setlength{\tabcolsep}{10pt} 
\renewcommand{\arraystretch}{1.7}
\begin{tabular}{|l||c|l|}
\hline
Quasiparticle property & Symbol & Relation to self energy \\
\hline
Energy & $E$ & $E={\rm Re}\left[\Sigma(0,E)\right]$\\
Residue & $Z$ & $Z=\left(1-\left.\pdv{{\rm Re}\left[\Sigma(0,\omega)\right]}{\omega}\right|_{\omega=E}\right)^{-1}$\\
Effective mass & $m^*$ & $m^*=\frac{m}{Z}\left(1+\left.\pdv{{\rm Re}\left[\Sigma(\p,E)\right]}{\ep}\right|_{p=0}\right)^{-1}$\\
Damping & $\Gamma$ & $\Gamma=-Z\,{\rm Im}\left[\Sigma(0,E)\right]$\\
Contact & $C$ & $\frac{C}{8\pi m_r}=\pdv{E}{(-1/a)}= Z\, \pdv{{\rm Re}\left[\Sigma(0,E)\right]}{(-1/a)}$ \\
Particles in dressing cloud & $\Delta N$ & $\Delta N=-\pdv{E}{\mu}=-Z\pdv{{\rm Re}\left[\Sigma(0,E)\right]}{\mu}$ \\
\hline
\end{tabular}
\vspace*{10pt}
\caption{Quasiparticle properties of an impurity in a medium. The quasiparticle energy $E$ is found as the solution of a transcendental equation, and serves as an input into the remaining expressions.
\label{tab:qp}}
\end{table*}

\subsection{Quasiparticle properties}

The interactions between the impurity and the medium lead to excitations of the medium, and consequently the state corresponding to the bare impurity on top of an unperturbed bath  is no longer an eigenstate of the system Hamiltonian. As illustrated in Fig.~\ref{fig:schematic}, the resulting impurity \textit{quasiparticle} has modified properties such as energy, mass, and residue (squared wave function overlap with the non-interacting state), and likewise the quasiparticle can acquire a finite lifetime. We now outline how these formally appear in the theory.

The quasiparticle properties at temperature $T=1/\beta$ are all encoded in the retarded impurity Green's function
\begin{align} \label{eq:green-t}
    {\cal G}(\p,t)=-i\Theta(t)\tr\left[\hat\rho_{\rm med} \hat c_\p(t)\hat c_\p^\dag(0)\right],
\end{align}
which is written in terms of the time-dependent impurity operator $\hat c_\p(t)=e^{i\hat Ht}\hat c_\p e^{-i\hat Ht}$.
Here, $\hat \rho_{\rm med}=e^{-\beta\hat H_{\rm med}}/\tr\left[e^{-\beta\hat H_{\rm med}}\right]$ is the medium density matrix and the trace is over the eigenstates of the medium in the absence of the impurity. In a time-independent system, it is convenient to introduce the impurity Green's function as a function of frequency via the Fourier transform
\begin{align}
    G(\p,\omega)=\int dt\, e^{i\omega t}{\cal G}(\p,t).
\end{align}
This satisfies the Dyson equation
\begin{align}
    G(\p,\omega) &=G_0(\p,\omega)+G_0(\p,\omega)\Sigma(\p,\omega)G(\p,\omega)  \notag \\
    &= \frac1{G_0(\p,\omega)^{-1}-\Sigma(\p,\omega)}, 
\end{align}
in terms of the impurity self energy $\Sigma$ and the bare impurity propagator $G_0(\p,\omega)=1/(\omega-\ep+i0)$, where the infinitesimal factor $+i0$ shifts the pole slightly into the lower half plane.

The impurity self energy allows us to extract the quasiparticle properties~\cite{FetterBook}. 
In particular, the presence of a quasiparticle is related to a pole of the Green's function, and in the vicinity of this pole we have
\begin{align} \label{eq:green-pole}
    G(\p,\omega)\simeq \frac{Z}{\omega-E-p^2/2m^*+i\Gamma}
\end{align}
for small momenta. Here, the energy $E$, residue $Z$, effective mass $m^*$, and damping rate $\Gamma$ are related to the self energy as outlined in Table~\ref{tab:qp}. The quasiparticle properties manifest themselves in the spectral function, defined from the Green's function as 
\begin{align}
\label{eq:Apomega}
    A(\p,\omega)=-\frac1\pi{\rm Im}\left[G(\p,\omega)\right],
\end{align}
which is one of the key experimental observables of quantum impurity physics (see Fig.~\ref{fig:schematic}).

In addition to the above properties, which are familiar from Fermi liquid theory, the quasiparticles are also characterized by quantities that encode the impurity-medium correlations. 
Of particular interest is the impurity Tan contact~\cite{Tan2008a,Tan2008b},
\begin{align}
    C=8\pi m_r \, \pdv{E}{(-1/a)},
\end{align}
which is related to the probability of a medium particle being close to the impurity. This contact governs the occupation at large momenta~\cite{Tan2008a} and, when the polaron energy $E$ corresponds to the ground-state energy at zero temperature, it is a thermodynamic quantity that plays a role in various thermodynamic properties~\cite{Tan2008a,Braaten2008}.
Note that the impurity contact can also be a thermodynamic variable at finite temperature, but in this case it should strictly speaking be defined from the free energy rather than the polaron energy~\cite{Liu2020prl}.

Another quantity of interest is the number of bath particles in the impurity dressing cloud \cite{Massignan2005} 
\beq\label{DeltaN}
\Delta N=-\frac{\partial E}{\partial \mu},
\eeq
defined as the number of particles that must be added to the medium in order to keep its chemical potential (i.e., the medium density far away from the impurity) fixed when the impurity is inserted into the system.

The energy, the contact and the number of particles in the dressing cloud are actually tightly linked, as can be shown by a simple argument based on dimensional analysis~\cite{FetterBook,Werner2008}. 
Since the polaron energy is independent of the chosen unit of length, we must have $E(T,a,n,R)=\lambda^{-2}E(T\lambda^{2},a/\lambda,n\lambda^{3},R/\lambda)$ for arbitrary scaling factor $\lambda>0$. Here, $R$ represents an extra length scale which may influence the energy, such as $a_B$ in the case of Bose polarons or an effective range of the resonance. Taking $dE/d\lambda=0$ and setting $\lambda=1$, we obtain
\beq
E = \left(T\partial_T-\frac{a}{2}\partial_a+\frac{3n}{2}\partial_n-\frac{R}{2}\partial_{R}\right)E.
\eeq 
For a broad Feshbach resonance at $T=0$, this yields
\begin{subequations} \label{eq:DeltaN2}
\begin{align} 
    \Delta N &= -\frac1{E_F}\left[\frac C{16\pi m_ra}+E\right] \hspace{3mm} {\rm (Fermi)} \\
    \Delta N &=-\frac2{3\mu}\left[\frac C{16\pi m_ra}+E + \frac{a_B}{2} \pdv{E}{a_B} 
    \right] \hspace{3mm} {\rm (Bose)}
\end{align}
\end{subequations} 
in the case of Fermi \cite{Massignan2012} and Bose polarons. These expressions differ due to the additional length scale $a_B$ in the Bose gas and the density scaling of each chemical potential, which is $\mu=E_F$ and $\mu=4\pi a_B n/m_{\rm med}$ for Fermi and Bose media, respectively. Moreover, the additional term in the Bose case can be regarded as a three-body contact~\cite{Braaten2011} involving the impurity and two bosons \footnote{Since the derivative of energy with respect to the Bose-Bose scattering length acts on the impurity energy, this term necessarily involves the impurity and two bosons.}.
As a direct consequence of Eq.~\eqref{eq:DeltaN2}, the energy of attractive Fermi polarons at the unitary point (where $a$ diverges) satisfies $\Delta N = -E/E_F$.
In general, we define $C$ and $\Delta N$ from the impurity self energy, as displayed in Table~\ref{tab:qp}, such that we can apply these relationships to arbitrary quasiparticles with a finite lifetime---e.g., the repulsive polaron---not just those that correspond to an eigenstate.

\begin{table*}[t]
\centering
\setlength{\tabcolsep}{10pt} 
\renewcommand{\arraystretch}{1.7}
\begin{tabular}{|c||c|c|}
\hline
& Fermi polaron & Bose polaron \\
\hline
$E$ & $\frac{4\pi n a}{m}\left(1+\frac3{2\pi}k_Fa\right)$ \cite{Bishop1973}
& $\frac{4\pi n a}{m}\left(1+\frac{8\sqrt2}{3\pi}\frac a\xi\right)$ \cite{Novikov2009} \\
$Z$ & $1-\frac2{\pi^2}(k_Fa)^2$ \cite{Trefzger2013} & $1-\frac{2\sqrt{2}}{3\pi}\frac{a^2}{a_B\xi}$ \cite{Christensen2015}\\
$m^*/m$ & $1+\frac2{3\pi^2}(k_Fa)^2$ \cite{Bishop1973}
& $1+\frac{16\sqrt2}{45\pi}\frac{a^2}{a_B\xi}$ \cite{Casteels2014} \\
$\Gamma$ & $\frac{8(k_Fa)^4}{9\pi^3}E_F$ \cite{Adlong2020} & $\frac{4\pi n a^2}{3m \xi}$, \,  if $a = a_B$ \\
$C$ & $16\pi^2 na^2\left(1+\frac3\pi k_Fa\right)$ & $16\pi^2 na^2\left(1+\frac{16\sqrt2}{3\pi}\frac a\xi\right)$ \\
$\Delta N$ & $-\frac2\pi k_Fa-\frac4{\pi^2}(k_Fa)^2$ \cite{Massignan2011} & 
$-\frac{a}{a_B}-\frac{4\sqrt2}{\pi}\frac{a^2}{a_B\xi}$
\\
\hline
\end{tabular}
\vspace*{10pt}
\caption{Perturbation theory results for quasiparticle properties in the case $m=m_{\rm med}$, evaluated up to 2$^{\rm nd}$ order in the impurity-medium scattering length. The result for the damping rate $\Gamma$ refers to the repulsive polaron in the case where we have short-range attractive impurity-medium interactions (see Section~\ref{sec:qp-lifetime}).
\label{tab:pt}}
\end{table*}

\subsubsection{Limit of weak interactions}

In the limit of weak impurity-medium interactions $k_F |a| \ll 1$, one can apply perturbation theory to obtain exact analytical expressions for the quasiparticle properties. Here, the scattering length $a$ can be either positive or negative, corresponding to repulsive or attractive polarons, respectively. Most notably, the behavior up to order $(k_F a)^2$ is universal and independent of the microscopic details of the interactions for both Fermi and Bose polarons. Indeed, the perturbative expressions at this order are even insensitive to whether the underlying interactions are attractive or repulsive.

Table~\ref{tab:pt} summarizes the perturbative results up to ${\cal O}(a^2)$ or to the lowest non-vanishing order for both types of polarons, where, for simplicity, we specialize to the case of an impurity of the same mass as the medium particles and we take $T$ to be smaller than the interaction energy shift such that it has no effect at this order. In the case of the Fermi polaron, the perturbative expansion was first carried out by Bishop~\cite{Bishop1973}. For the Bose polaron, the expansion was first considered by Novikov and Ovchinnikov~\cite{Novikov2009} and requires $|a|/\xi\ll1$ and $a^2/(a_B\xi)\ll1$, where we assume that the Bose medium is condensed with $\xi=1/\sqrt{8\pi n a_B}$ the condensate healing length (note that the Bose polaron expansion has been carried out to even higher order in $a$ ---for details, see Ref.~\cite{Christensen2015}). From Table~\ref{tab:pt} we see that, apart from $\Gamma$ and the number of particles in the dressing cloud, the leading order behavior is the same, i.e., the statistics of the medium only enters at higher order. However, already beyond leading order, there are intriguing differences between the two cases. While the expressions for the correction to, e.g., the energy and the contact look quite different, they take a similar form when formulated in terms of the compressibility of the two media. For instance, relating the 2nd order term to the respective speeds of sound, $c_F=k_F/(\sqrt3m)$ and $c_B=1/(\sqrt2m\xi)$, we find that the correction to the energy is comparable: $\frac3{2\pi}k_Fa\approx 0.8mac_F$ and $\frac{8\sqrt2}{3\pi}\frac a\xi\approx1.7mac_B$.

\subsubsection{Strong-coupling polarons}

Going beyond the limit of weak impurity-medium interactions, the polaron problem generally becomes analytically intractable and one must turn to approximation methods. In particular, the metastable repulsive branch in the strong-coupling regime where $|k_Fa|\gtrsim1$ has been tackled with a wide variety of methods. For the Fermi polaron, these include variational methods~\cite{Cui2010,Parish2016,Liu2019,Adlong2020}, 
diagrammatic Monte Carlo~\cite{Goulko2016},
diagrammatic methods ($T$-matrix, many-body $T$-matrix and beyond)~\cite{Massignan2011,Tajima2018, Mulkerin2019,Tajima2021,Hu2022}, 
virial expansion~\cite{Mulkerin2019}, functional renormalization group~\cite{Schmidt2011}, and functional determinants \cite{Schmidt2018}. 
In the case of the Bose polaron, the repulsive branch has been investigated using diagrammatic methods~\cite{Rath2013,Guenther2018}, variational methods~\cite{Li2014,Jorgensen2016,Shchadilova2016,Field2020}, quantum Monte Carlo~\cite{Ardila2015,Ardila2018,Ardila2020}, virial expansion~\cite{Sun2017}, and functional renormalization group~\cite{Grusdt2018,Isaule2021}.
We will briefly discuss the variational method in Section~\ref{sec:qp-lifetime}; however an in-depth discussion of the other methods is beyond the scope of this review.

\section{Experimental probes} \label{sec:expt}

\begin{figure*}[t]
\begin{center}
\includegraphics[scale=1.15]{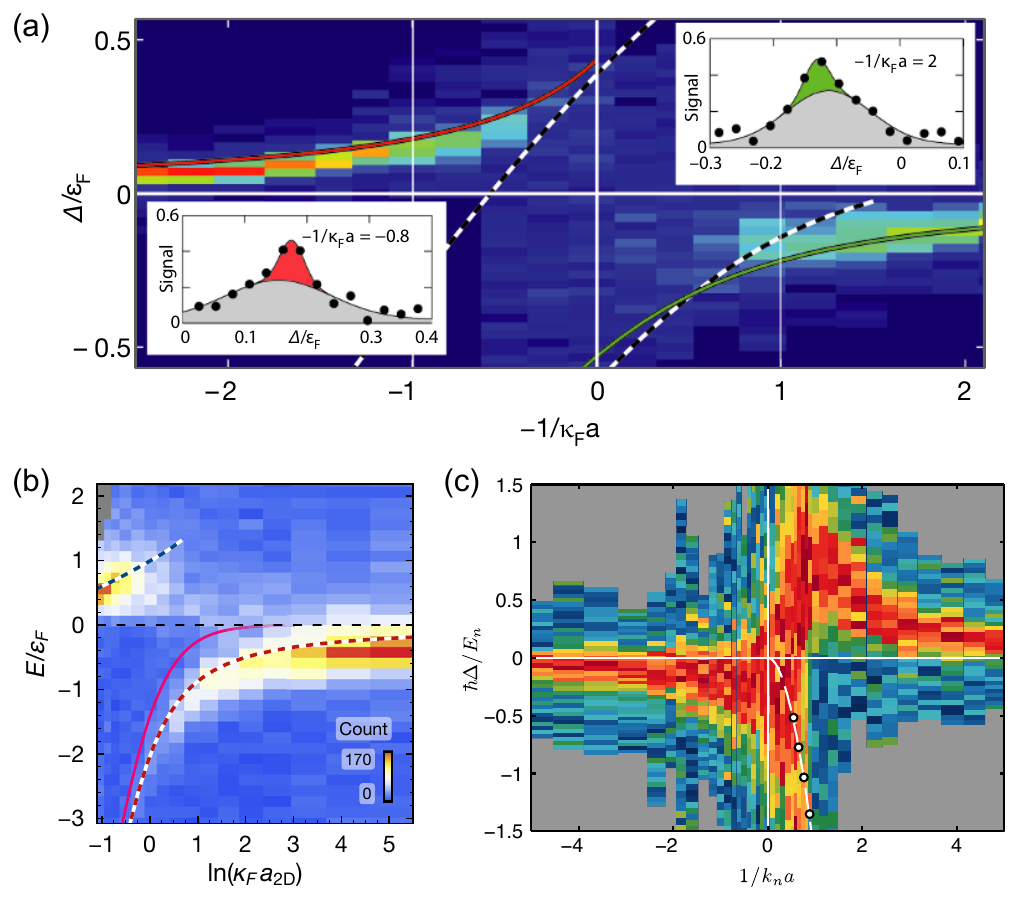}
\caption{Experimental RF injection spectroscopy of impurities immersed in different media: (a) a three-dimensional (3D) Fermi gas \cite{Kohstall2012}, (b) a two-dimensional (2D) Fermi gas \cite{Oppong2019}, and (c) a 3D Bose-Einstein condensate \cite{Jorgensen2016}. For the 3D and 2D fermionic backgrounds, the impurity-medium interaction strength is encoded by the dimensionless parameters $1/k_Fa$ and $\log(k_Fa)$, and are tuned via a narrow Feshbach resonance \cite{Kohstall2012} and an orbital Feshbach resonance \cite{Oppong2019}, respectively. 
The vertical energy scale is normalized to $E_F$ and the zero corresponds to the frequency of the atomic RF transition in the absence of the medium. The repulsive polaron energy branch is clearly visible in all cases at positive RF detuning, ceasing to be well defined upon approaching unitarity-limited interactions from the repulsive side, i.e.~$a>0$ or $\ln(k_F a_{2D})<0$. Panel (a) is adapted from Ref.~\citenum{Kohstall2012}, panel (b) from Ref.~\citenum{Oppong2019}, and panel (c) from Ref.~\citenum{Jorgensen2016}.}
\label{fig:FermiRFexp}
\end{center}
\end{figure*}

We now provide a brief description of the techniques used to probe and characterize quasiparticle properties in current experiments with ultracold atoms, and we discuss how the available experimental measurements compare with existing theories. This short overview focuses mainly on the metastable repulsive polaron state. 

The most well-established experimental protocol for probing quasiparticles in ultracold atomic mixtures is radio-frequency (RF) spectroscopy \cite{Torma2016}. 
In the last decade, RF spectroscopy has been the technique of choice to precisely address a variety of properties in ultracold atom experiments \cite{Vale2021}. It has permitted the first direct observation of both attractive \cite{Schirotzek2009} and repulsive  \cite{Kohstall2012,Koschorreck2012} Fermi polarons, and since then it has been exploited extensively in both Fermi and Bose polaron studies \cite{Zhang2012,Scazza2017,Yan2019a,Hu2016,Jorgensen2016,Yan2019,Fritsche2021}. 
RF spectroscopy involves two internal (hyperfine) states of the impurity atoms which are coupled by an oscillating RF field. Because of its long wavelength, the RF field is essentially uniform over the sub-millimeter scale of atomic samples, and RF photon absorption transfers a negligible momentum to the atoms. 
The two coupled impurity states are chosen so as to feature different interaction strengths with the surrounding medium, which may in turn be composed of a third hyperfine state of the same atomic species or an entirely different atomic species. 

In the case of the so-called \emph{injection} spectroscopy, a weak RF pulse transfers impurities from a (nearly) non-interacting state to another state featuring strong interactions with the medium particles. To perform spectroscopy at varying interaction strength, the impurity-medium scattering length $a$ is tuned by means of a Feshbach resonance \cite{Chin2010}. 
Within the linear response regime, the RF signal is given by (see, e.g., Refs.~\citenum{Punk2007,Massignan2008c})
\begin{equation}
\label{eq:RFhom}
I_{\rm inj}(\omega)\propto
\sum_\p
n_\p\, A(\p,\epsilon_{\p}+\omega).
\end{equation}
Here, $n_\p$ is the initial momentum distribution function of the impurities, $A(\p,\omega)$ is the impurity spectral function given by Eq.~\eqref{eq:Apomega}, and we measure the frequency $\omega$ from the bare transition frequency between the initial and final hyperfine states.
Since the RF pulse has a finite duration, in order to describe the experimental spectral response, Eq.~\eqref{eq:RFhom} has to be convoluted with a filter function whose width is inversely proportional to the duration of the RF pulse.
To maintain a direct correspondence between the experimentally recorded spectrum and the impurity spectral function, care must be taken to operate sufficiently close to the linear response regime, as well as to work with sufficiently low impurity concentrations and sample temperatures. Otherwise, for instance, the first moment of the experimental RF signal may significantly deviate from the quasiparticle energy \cite{Kohstall2012,Ardila2018}.

\begin{figure}[t]
\begin{center}
\includegraphics[scale=0.33]{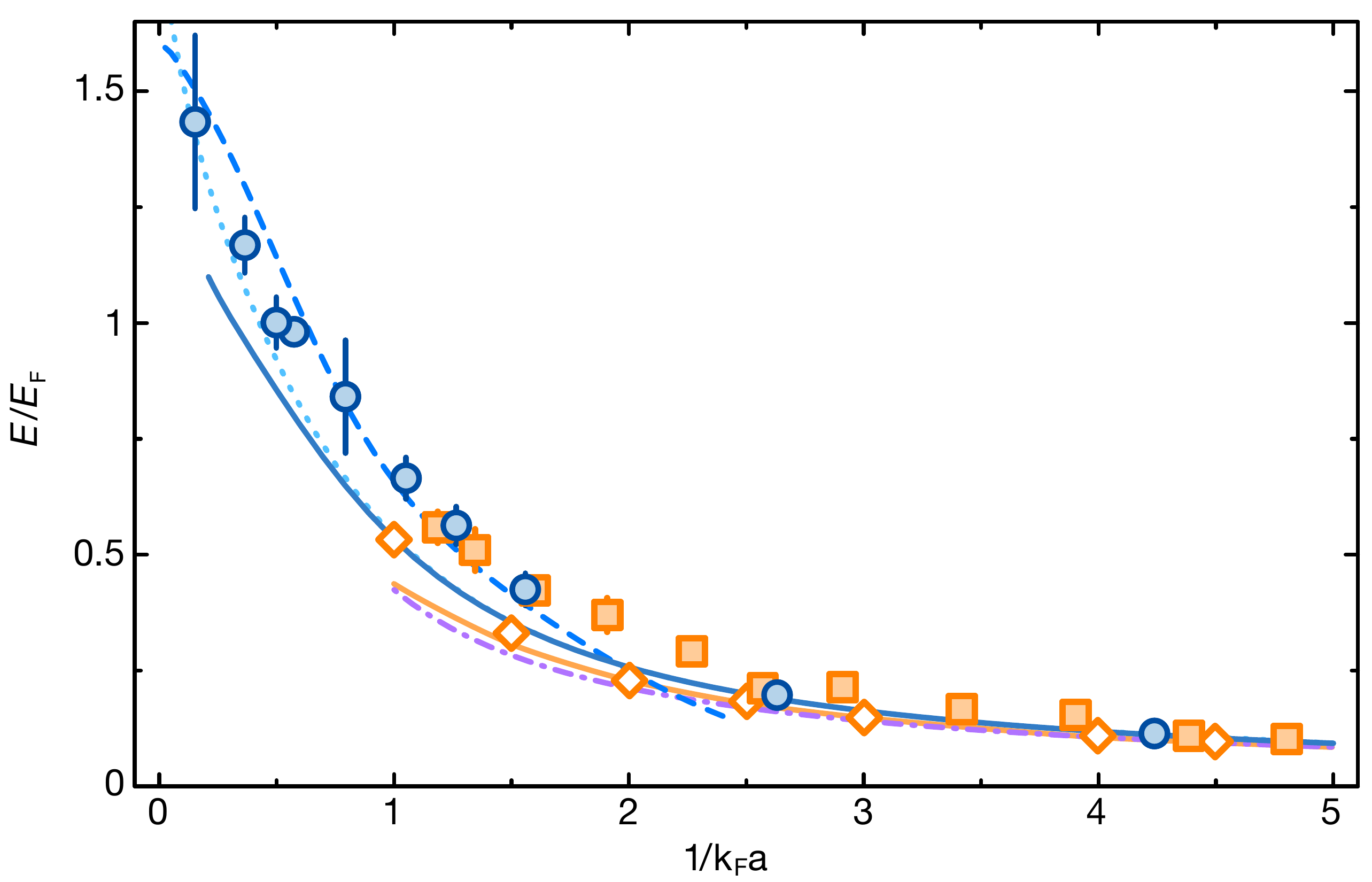}
\caption{Energy of repulsive Fermi polarons (blue symbols and lines) and Bose polarons (orange symbols and lines) in the equal-mass case. RF spectroscopy experimental data are taken from Ref.~\citenum{Scazza2017} (filled blue circles) and Ref.~\citenum{Ardila2018} (filled orange squares). We also display theoretical results obtained by a non self-consistent $T$-matrix approach \cite{Massignan2011} (solid blue line), functional renormalization group \cite{Schmidt2011} (dashed  blue line) and a variational wavefunction \cite{Cui2010} (dotted light blue line) for the repulsive Fermi polaron, and by the truncated basis method (TBM) \cite{Jorgensen2016} (solid orange line) and fixed-node diffusion QMC \cite{Ardila2018} (empty orange diamonds) for the repulsive Bose polaron. The mean-field prediction $E=4\pi n a/m \equiv E_F \frac{4}{3\pi} k_Fa$ (purple dot-dashed line) coincides for the Bose and Fermi polarons.}
\label{fig:PolEnergies}
\end{center}
\end{figure}

\begin{figure}[t]
\begin{center}
\vspace{-1mm}
\includegraphics[scale=0.96]{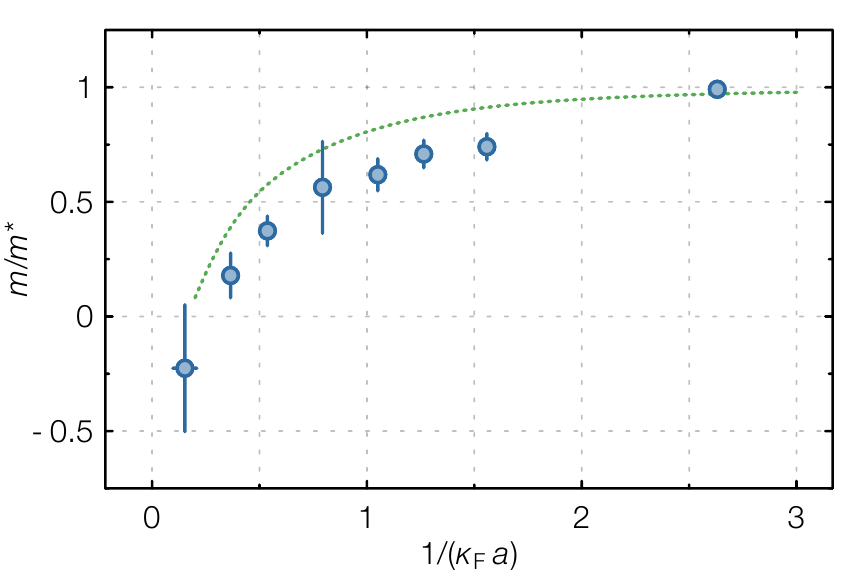}
\caption{Effective mass of the repulsive Fermi polaron in the equal-mass case: experimental results (blue circles) and theoretical results (dotted green line) from a $T$-matrix approach \cite{Massignan2011}. $m^*$ is observed to diverge when approaching the unitary limit. The remaining deviation between theory and experimental data may result from higher-order particle-hole excitations not taken into account in the calculation, as well as from weak polaron-polaron interactions which are neglected in the experimental extraction of $m/m$*. The figure is adapted from Ref.~\citenum{Scazza2017}.}
\label{fig:PolMass}
\end{center}
\end{figure}

Injection spectroscopy allows one to probe the entire many-body spectrum of strongly interacting impurities, since it addresses both the ground and the excited states of the impurity-medium Hamiltonian [see Eq.~\eqref{eq:Hgen}]. Indeed, repulsive Fermi polarons have been revealed by injection spectroscopy (see Fig.~\ref{fig:FermiRFexp}), first in a highly imbalanced, heteronuclear Fermi mixture \cite{Kohstall2012} and subsequently in a homonuclear spin mixture \cite{Scazza2017}. Similarly, injection RF spectroscopy of mobile impurities immersed in a Bose-Einstein condensate has allowed the observation of repulsive Bose polarons \cite{Hu2016, Jorgensen2016}.
This RF spectroscopy technique has been recently extended to the optical domain \cite{Oppong2019}, exploiting the clock transition and tunable clock state interactions in alkaline-earth-like atoms \cite{Scazza2014,Hofer2015}.

The opposite protocol, where a strongly interacting state is flipped into a non-interacting state, is termed \emph{ejection} spectroscopy, and has been extensively used to probe the ground state of strongly interacting mixtures with arbitrary population imbalance \cite{Vale2021}. While ejection spectroscopy in general depends on the initial occupation of states in the strongly interacting system, it simplifies in the case of a very low impurity concentration, where the impurities are uncorrelated and the distribution function in Eq.~\eqref{eq:RFhom} reduces to a Boltzmann distribution. There, the ejection and injection spectra are directly related via~\cite{Liu2020}
\begin{align}
    I_{\rm ej}(\omega)=e^{\beta\omega}e^{\beta\Delta F}I_{\rm inj}(-\omega),
\end{align}
in terms of the difference in free energy between the interacting and the non-interacting impurity, $\Delta F$. The exponential prefactor suppresses the repulsive branch at positive energies, which clearly illustrates why ejection spectroscopy is ideally suited to investigations of ground-state properties.

\begin{figure*}[t]
\begin{center}
\includegraphics[scale=0.83]{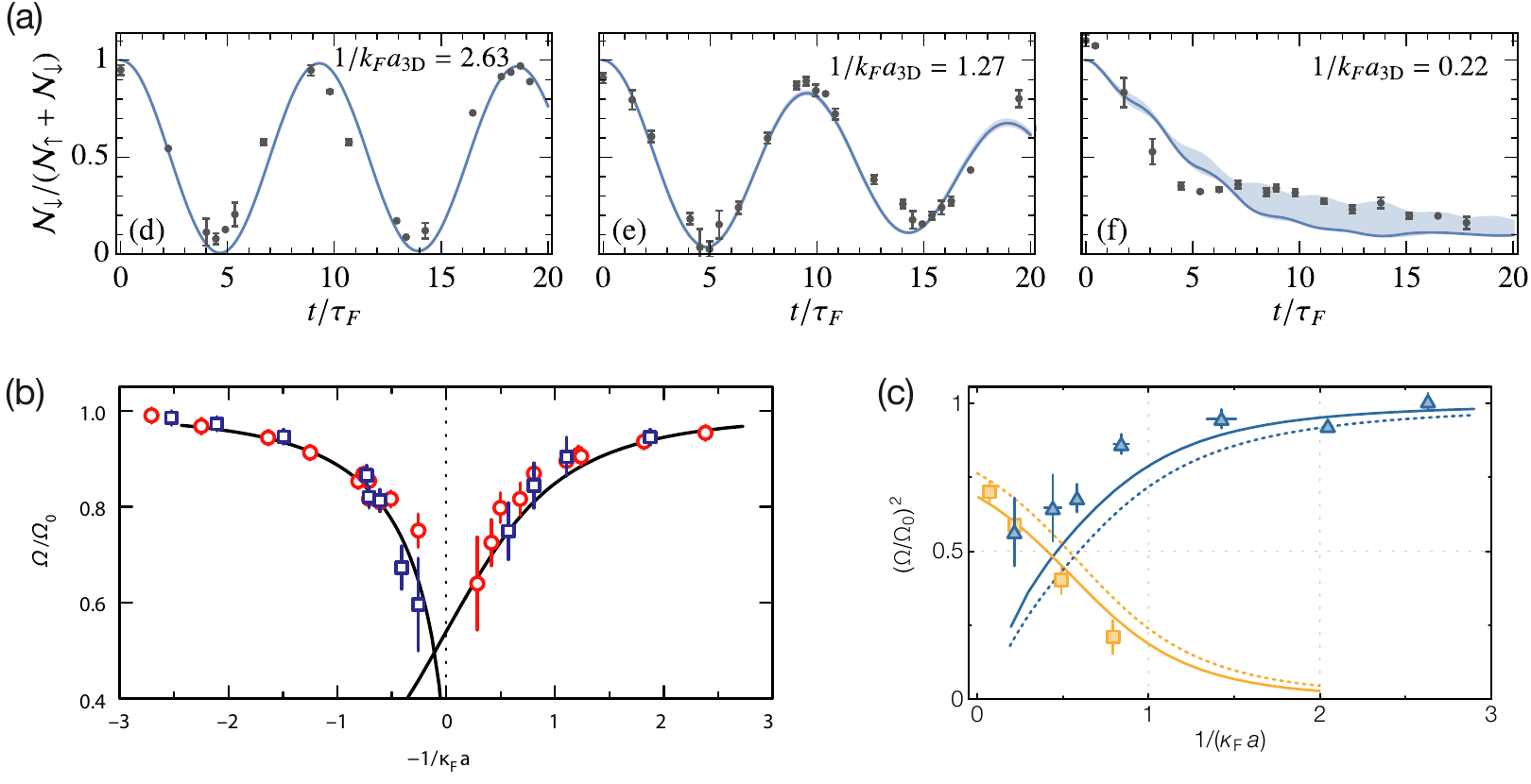}
\caption{Rabi oscillations between non-interacting and strongly interacting impurity states in a fermionic medium. (a) Points are experimental measurements, while lines denote theoretical predictions based on a variational ansatz with no free parameters \cite{Adlong2020}. Shaded regions allow for experimental parameter uncertainties. (b)-(c) Normalized Rabi frequency extracted from experimental Rabi oscillations (symbols) in mass-imbalanced [in (b)] \cite{Kohstall2012} and equal-mass [in (c)] \cite{Scazza2017} mixtures are compared to theoretical predictions for $\sqrt{Z}$ [in (b)] and $Z$ [in (c)] (lines), obtained with a $T$-matrix method \cite{Massignan2011,Kohstall2012,Scazza2017}. Solid lines in panel (c) take into account the effect of small initial-state interactions. 
Panel (a) is adapted from Ref.~\citenum{Adlong2020}, 
panel (b) from Ref.~\citenum{Kohstall2012}, 
and panel (c) from Ref.~\citenum{Scazza2017}.}
\label{fig:PolRabi}
\end{center}
\vspace*{-10pt}
\end{figure*}

The impurity RF response encapsulates a variety of essential information about polaronic states \cite{Massignan_Zaccanti_Bruun}. Most importantly, when the spectrum contains a well-defined quasiparticle, i.e~a coherent excitation of the medium, the main contribution to the spectral function stems from the quasiparticle pole [see Eq.~\eqref{eq:green-pole}], and thus the mean and the width of the RF signal $I_{\rm inj}(\omega)$ yield a measure of the mean quasiparticle energy $E + \langle p^2\rangle/2m^*$ and quasiparticle width $\Gamma$, respectively (see also the inset of Fig.~\ref{fig:schematic}). Ejection spectroscopy additionally allows one to extract the contact $C$ of the attractive polaron \cite{Yan2019a,Yan2019,Liu2020prl,Ness2020,Vale2021}, which is encoded in the high-frequency tail of the RF signal \cite{Braaten2010}. 
RF spectra have been successfully employed to extract the quasiparticle energies of both attractive and repulsive (Bose and Fermi) polarons as a function of the impurity-medium interaction strength \cite{Schirotzek2009,Kohstall2012,Koschorreck2012,Jorgensen2016,Hu2016,Scazza2017,Fritsche2021} and temperature \cite{Yan2019a,Yan2019}. Experimental results for the energies of repulsive Fermi and Bose polarons at the lowest achieved temperatures in homonuclear mixtures are shown in Fig.~\ref{fig:PolEnergies}, where they are also compared to existing theories.
Note that, while the energy of Fermi polarons depends only on a single parameter (up to second order, only on $k_Fa$), the energy of Bose polarons depends additionally on $a/\xi$ \cite{Guenther2021}. 
Future spectroscopic studies in homogeneous Bose and Fermi gases could resolve the remaining discrepancies between available measurements performed with harmonically trapped samples and calculations at fixed density.

The quasiparticle width can also be obtained from the RF spectra within linear response, although a precise extraction becomes increasingly difficult 
as the quasiparticle spectral peak becomes either narrow or broad with respect to the polaron energy.
Here, there are strong distinctions between the attractive and repulsive branches, and between Fermi and Bose polarons. In the latter Bose polaron case, the attractive polaron quasiparticle with a finite residue remains the ground state at all interaction strengths, and hence the quasiparticle width $\Gamma$ always vanishes at zero temperature. However, for the Fermi polaron, the attractive polaron is the many-body ground state only for $1/k_Fa$ up to the so-called polaron-molecule transition \cite{Prokofev2008,Punk2009,Combescot2009}. At finite temperature, the attractive polaron width in the RF signal is simply linked to the impurity-medium collision rate \cite{Yan2019a,Bruun2008}.
Conversely, the metastable repulsive polaron retains a finite lifetime even for a zero-momentum impurity, which for weak coupling is governed by a many-body dephasing mechanism rather than by the decay to lower-lying attractive states (see below and Section \ref{sec:qp-lifetime}).

The polaron effective mass is rather challenging to probe, as it requires one to access the polaron dispersion relation, i.e., to probe the spectral response at different impurity momenta [see Eq.~\eqref{eq:green-pole}]. An effective spectroscopic technique relies on the fact that, due to the enhanced effective mass of polarons $m^* > m$, a moving polaron transferred into a final non-interacting state, or vice versa, will have a RF response peak that varies with the impurity momentum. 
By varying the concentration of a fermionic minority component -- still within the highly polarized limit -- it is possible to effectively vary the mean impurity momentum due to Pauli exclusion. This effect has been exploited both in ejection and injection RF spectroscopy to yield estimates of the attractive and repulsive Fermi polaron effective mass \cite{Schirotzek2009,Scazza2017} (see Fig.~\ref{fig:PolMass}). However, it is \textit{a priori}
difficult to quantify the impact of weak polaron-polaron interactions 
on such measurements at variable impurity concentration. Momentum-resolved ejection spectroscopy has also been exploited in a highly polarized spin mixture in 2D to directly measure the impurity energy dispersion as a function of momentum, yielding an estimate of the attractive Fermi polaron mass \cite{Koschorreck2012}. Experimental measurements of the Bose polaron effective mass are presently lacking; for this, novel Raman spectroscopic techniques could be exploited to impart a controlled momentum to impurities \cite{Ness2020}.

\begin{figure*}[t]
\begin{center}
\includegraphics[scale=0.515]{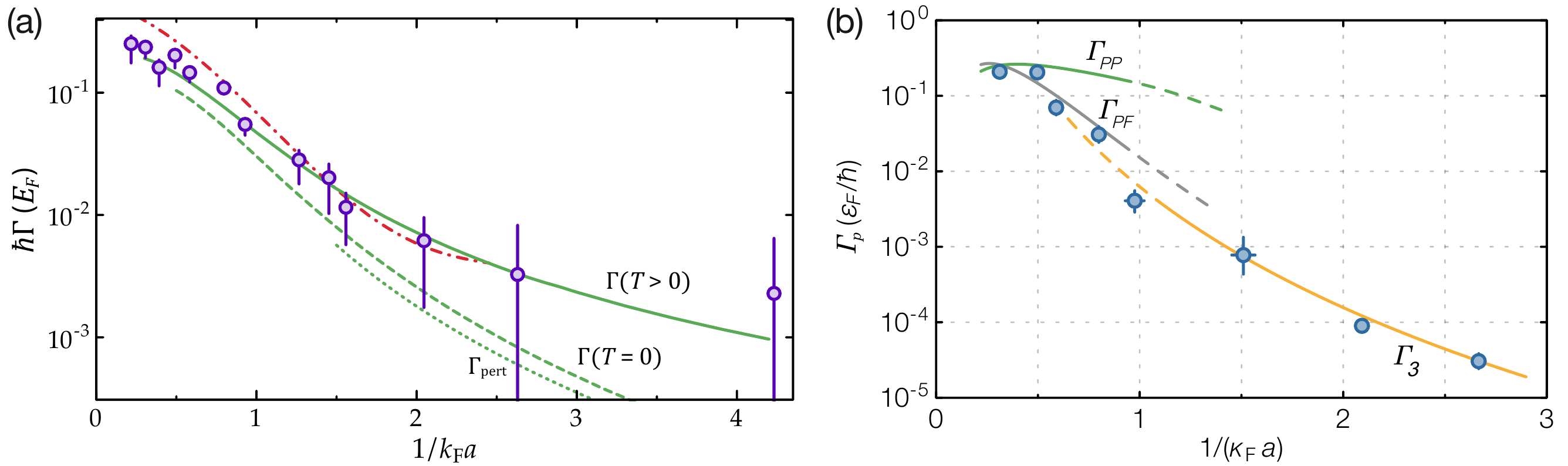}
\caption{
Stability of the repulsive Fermi polaron in the equal-mass, broad-resonance case.  (a)~Decoherence rate $\Gamma$ of the repulsive Fermi polaron. (b) Population decay rate $\Gamma_p$ of the repulsive branch to lower-lying states. 
Circles are experimental data points from Ref.~\citenum{Scazza2017}, obtained by measuring: (a) the damping rate of Rabi oscillations on the repulsive polaron, or (b) the relaxation of the upper branch measured via a double-pulse experiment. 
In panel (a), lines denote the quasiparticle spectral width $\Gamma$ calculated with a variational method \cite{Liu2019,Liu2020,Adlong2020} at $T=0$ (dashed, green) and $T=0.13\,E_F$ (solid, green), with functional renormalization group (red, dot-dashed) \cite{Schmidt2011}, and at the first non-vanishing order in perturbation theory, $\Gamma_\mathrm{pert}=\frac{8(k_Fa)^4}{9\pi^3}E_F$ \cite{Adlong2020} (dotted, green -- see Table \ref{tab:pt}). In panel (b), lines denote the upper branch decay rate predicted by considering polaron-to-polaron recombination (green) \cite{Massignan2011}, polaron-to-free particle recombination (grey) \cite{Scazza2017}, and three-body molecular recombination (yellow) \cite{Petrov2003}.
 Panel (b) is adapted from Ref.~\citenum{Scazza2017}.}
\label{fig:FermiStability}
\end{center}
\end{figure*}

The injection and ejection RF spectroscopic probes which we have considered thus far are suited to precisely access quasiparticle spectral properties, as they are well resolved in energy. However, they cannot be exploited to track the formation of quasiparticles in real time, because they lack the time resolution necessary to monitor the rapid build-up of the polaron dressing cloud. Conversely, dynamical probes such as many-body Ramsey interferometry are well adapted to this scope~\cite{Goold2011,Knap2012,Parish2016,Schmidt2018,Liu2019}, and have indeed enabled the experimental observation of the formation (and the decoherence) of both Fermi and Bose polarons~\cite{Cetina2016,Skou2021}, with the long-time value of the Ramsey contrast being connected to the quasiparticle residue \cite{Cetina2016,Parish2016}. It has also been shown that coherently driving Rabi oscillations between the two impurity states involved in the RF spectroscopy protocol is a powerful complementary technique to probe the polaron quasiparticle properties \cite{Parish2016,Adlong2020}. In particular, the residue $Z$ is directly connected to the Rabi frequency $\Omega$ normalized to the bare Rabi frequency $\Omega_0$ \cite{Kohstall2012}, which is obtained by performing Rabi oscillations in the absence of the medium. Indeed, $\sqrt{Z}$ essentially quantifies the wavefunction overlap between the polaron state and the bare, non-interacting impurity, and one finds that $Z \simeq (\Omega^2+\Gamma^2)/\Omega_0^2$ as long as $\Gamma\lesssim\sqrt{Z}\Omega_0$~\cite{Adlong2020}. This connection affords an alternative route to estimating the quasiparticle residue from the area of the coherent spectral response \cite{Schirotzek2009,Yan2019a,Yan2019}, a method which is extremely challenging once the coherent quasiparticle resonance approaches the onset of the molecular continuum. 

Rabi oscillations have 
allowed experiments to obtain the quasiparticle residue of attractive and repulsive Fermi polarons down to weak couplings $|1/k_Fa| > 1$ \cite{Kohstall2012,Scazza2017,Oppong2019} (see Fig.~\ref{fig:PolRabi}), although the extension to a bosonic medium is currently lacking. Moreover, it has recently been shown theoretically that the damping rate of Rabi oscillations can be essentially identified with the Fermi polaron quasiparticle damping rate $\Gamma$ for coherent quasiparticles with $\Gamma \lesssim \sqrt{Z}\Omega_0$~\cite{Adlong2020}. 
Thus, if the repulsive branch is both present and remains coherent, 
valuable information on the lifetime of repulsive polarons may be extracted from the damping rate of Rabi oscillations in experiment. 
As shown in panel (a) of Fig.~\ref{fig:FermiStability}, the measured damping rate of Rabi oscillations of impurities in a Fermi gas is found to be in excellent agreement with the quasiparticle damping rate $\Gamma$ obtained from a finite-temperature variational calculation containing a single particle-hole excitation \cite{Liu2019,Liu2020,Adlong2020} --- see Section~\ref{sec:qp-lifetime} for a detailed discussion of this approach.

In order to quantify the stability of the repulsive branch against relaxation to lower-lying attractive states, double-pulse sequences have been employed to directly address the repulsive branch de-population rate $\Gamma_p$ \cite{Kohstall2012,Koschorreck2012,Scazza2017,Oppong2019,Fritsche2021} [see Fig.~\ref{fig:FermiStability}(b)]. This technique is weakly sensitive to momentum, because it requires fast $\pi$-pulses with large Rabi frequency $\Omega \sim E_F$, thus coupling to all impurities in the medium irrespective of their kinetic energy. Theoretical calculations of the upper branch population decay rate $\Gamma_p$, based on variants of $T$-matrix techniques including two and three particle processes, match well with experimental measurements in Fermi-Fermi and Fermi-Bose mixtures at a narrow Feshbach resonance with heavy impurities for all couplings \cite{Kohstall2012,Fritsche2021} and in the strongly-interacting region at a broad resonance in the mass-balanced case \cite{Scazza2017}. For the equal-mass, broad-resonance case \cite{Scazza2017} at weaker couplings $1/k_Fa \gtrsim 1$ the measured repulsive branch decay rate was instead found to match well with the
predicted rate of recombination into dimers via three-body processes, $\Gamma_3$ \cite{Petrov2003}, as expected in the spin-balanced regime of a Fermi mixture \cite{Sanner2012,Amico2018}. Future theory approaches beyond the one particle-hole approximation for Fermi polarons may succeed to describe the trend of $\Gamma_p$ for an extended range of couplings in the equal-mass, universal broad-resonance case.

To conclude, in a fermionic medium with $1/k_Fa \gtrsim  1$, the repulsive branch decay rate $\Gamma_p$ is generally significantly smaller than the quasiparticle width $\Gamma$ (see Fig.~\ref{fig:FermiStability}), in both mass-imbalanced and mass-balanced scenarios \cite{Kohstall2012,Scazza2017}, thereby showing that the lifetime of repulsive Fermi polarons is mostly limited by decoherence processes (see also Section \ref{sec:qp-lifetime}).
Future experiments with repulsive fermionic impurities immersed in a Bose medium, possibly realized by mass-imbalanced mixtures with reduced three-body losses \cite{Chen2021}, could further address the stability of repulsive Bose polarons.

\section{Repulsive quasiparticle stability} \label{sec:qp-lifetime}

\begin{figure*}[t]
\begin{center}
\includegraphics[scale=0.35]{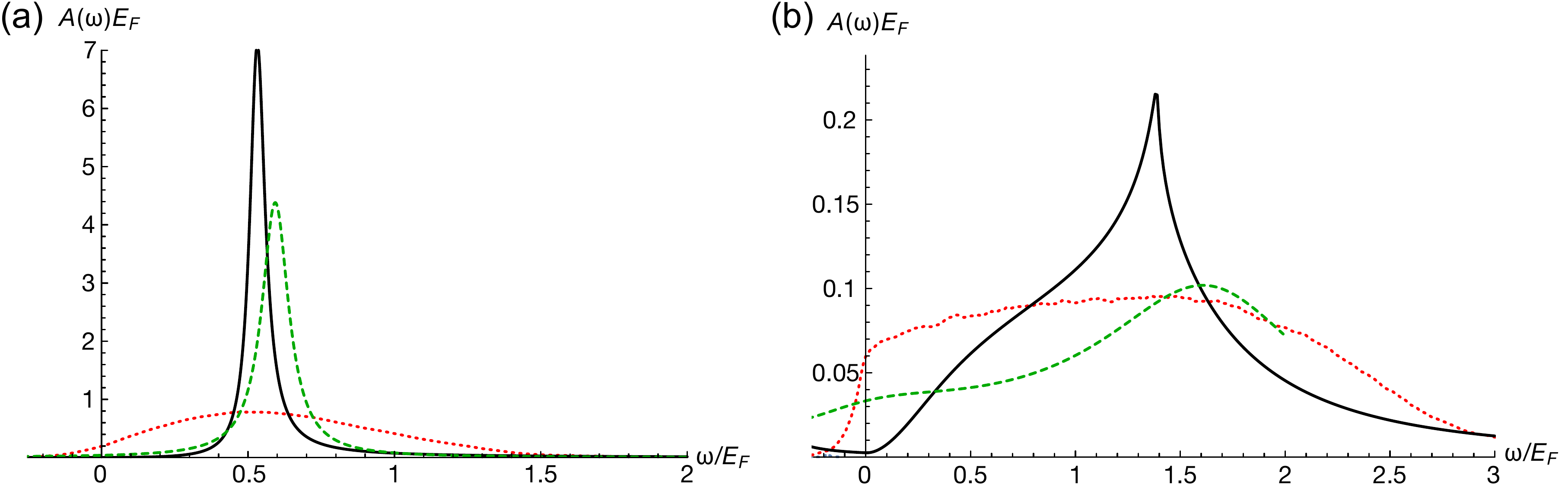}
\caption{Spectral function $A(\omega)\equiv A(\0,\omega)$ of the Fermi polaron for $m=m_{\rm med}$ and T=0, calculated using a $T$-matrix formalism \cite{Massignan2011} (black, solid), functional renormalization group \cite{Schmidt2011} (green, dashed), and diagrammatic Monte Carlo~\cite{Goulko2016} (red, dotted), where in the latter case, we display the maximally smooth solution~\cite{Goulko2016}. We show the results for slightly different values of $1/k_Fa\approx 1$ (left), and for $1/k_Fa=0$ (right).}
\label{fig:Fermispectrum}
\end{center}
\vspace*{-10pt}
\end{figure*}

The repulsive polaron quasiparticle is challenging to investigate theoretically, since it is a metastable object in quantum gases where the underlying short-range interactions are attractive.
In particular, there has been much debate about what controls the stability of the quasiparticle, as encoded by the quasiparticle damping rate 
$\Gamma$ in Eq.~\eqref{eq:green-pole}~\cite{Massignan2011,Schmidt2011,Kohstall2012,Cetina2015,Scazza2017,Adlong2020}. 
As discussed in Section~\ref{sec:expt}, $\Gamma$ directly manifests itself in the damping of Rabi oscillations and in the broadening of the quasiparticle peak in the spectral function. 
However, the quasiparticle damping rate has turned out to be far from trivial to calculate even in the weak-coupling limit~\cite{Scazza2017,Adlong2020}.

In this section, we will focus on the case of an impurity at rest, such that we can exclude any spectral broadening due to momentum relaxation~\cite{Bruun2008,Cetina2015}, where the polaron lowers its momentum via collisions. To gain insight into the quasiparticle stability, 
let us start with the non-interacting impurity-medium system at $T=0$ and then imagine gradually increasing the scattering length $a$ from zero, such that we adiabatically populate the repulsive branch. In the weak-coupling limit $k_F a \ll 1$ (Table~\ref{tab:pt}), the behavior of the metastable repulsive polaron  up to order $a^2$ is indistinguishable from that of the purely repulsive case, resembling an infinitely long-lived quasiparticle in the ground state. However, as we further increase the strength of the interactions, the metastable polaron develops a non-zero $\Gamma$, in contrast to the purely repulsive ground-state polaron. 
If we assume that $\Gamma$ is dictated by the decay into the attractive branch at negative energies (as it was initially assumed~\cite{Massignan2011,Kohstall2012}), then we can estimate its leading order behavior in $a$ from calculations of three-body recombination involving the impurity and two medium particles. 
This yields the scaling $a^6$ for the Fermi polaron case~\cite{Petrov2003} and $a^4$ for the Bose polaron when $a=a_B$~\cite{Braaten2006}, which can be understood from the fact that the population decay rate $\Gamma_p$ in this limit must scale as $n^2$ in the Bose case and $E_F n^2$ in the Fermi polaron case. 
However, an alternative mechanism for the quasiparticle damping has recently been proposed~\cite{Adlong2020}, whereby the repulsive polaron loses its coherence by coupling to the many-body continuum at positive energies, as illustrated in Fig.~\ref{fig:polaron-difference}. Crucially, this \textit{many-body dephasing} enters at a lower power of $a$ (see Table~\ref{tab:pt}) and thus dominates $\Gamma$ at weak coupling. This is consistent with experimental observations in Fig.~\ref{fig:FermiStability}, as discussed in Section \ref{sec:expt}.

As we approach strong interactions, we eventually expect both many-body dephasing and relaxation to the lower branch to determine the repulsive polaron lifetime. In this regime, the repulsive polaron becomes increasingly ill-defined and it is quite challenging to calculate the associated quasiparticle properties. 
Indeed, as illustrated in Fig.~\ref{fig:Fermispectrum}, different state-of-the-art calculations give inconsistent results for the spectral function in the case of the Fermi polaron, clearly indicating the need for future work. In particular, we see that the theories even disagree on the shape of the repulsive polaron peak at unitary, as well as yielding different results for the peak position and width.

In the following, we will delve further into one particular method, namely the variational method, since this describes many-body dephasing within the upper branch and provides a particularly intuitive way of accounting for excitations of the medium.
We will also show how the case of purely repulsive interactions results in a ground-state polaron with $\Gamma=0$ in this approach. Finally, we will examine the special case of an infinitely heavy impurity, since it admits an exact solution that serves as a useful reference point for the repulsive polaron, even though the quasiparticle residue $Z$ vanishes in this case.

\subsection{Variational description of attractive and repulsive polarons}

The basic idea of the variational method is to construct a variational form of the time-dependent impurity operator and then use this to approximate the impurity Green's function \eqref{eq:green-t}. In particular, it has proved remarkably fruitful to consider ``truncated'' operators with up to one or two excitations of the medium~\cite{Liu2019,Liu2020,Field2020,Adlong2020}. For the Fermi polaron, the simplest form we can write down for the zero-momentum impurity operator is~\cite{Liu2019}
\begin{align} \label{eq:TBM}
        \hat{c}_0(t) & \simeq \alpha_0(t) \, \hat{c}_0 
        + \sum_{\k, \q} \alpha_{\k \q}(t) \, \hat{f}^\dag_{\q} \hat{f}_{\k} \hat{c}_{\q-\k},  
\end{align}
where the time-dependent complex functions $\alpha_j(t)$ are variational parameters. Equation~\eqref{eq:TBM} can be regarded as the operator version of the variational wave function originally introduced by F. Chevy for the attractive Fermi polaron~\cite{Chevy2006}. However, unlike the original Chevy ansatz which only describes the attractive polaron ground state, the time-dependent variational operator can capture all the interacting impurity states at arbitrary temperature. In particular, the variational operator becomes exact in the high-temperature limit, where it becomes equivalent to the leading order contribution to the virial expansion~\cite{Liu2019}.

Similarly, the operator in the case of the Bose polaron has the form~\cite{Field2020}
\begin{align} \label{eq:TBM2}
        \hat{c}_0(t) & \simeq \alpha_0(t) \, \hat{c}_0 
        + \sum_{\q} \alpha^\q(t) \, \hat{b}^\dag_{\q} \hat{c}_{\q}
        + \sum_\k \alpha_{\k}(t) \, \hat{b}_{\k} \hat{c}_{-\k}  \notag\\
        &+ \sum_{\k, \q} \alpha_{\k}^\q(t) \, \hat{b}^\dag_{\q} \hat{b}_{\k} \hat{c}_{\q-\k} 
         + \dots,
\end{align}
where we have only explicitly written the terms with up to one excitation of the medium. 
Note that this differs slightly from the Fermi polaron case, since the second and third terms in Eq.~\eqref{eq:TBM2} describe bosons being scattered into and out of the condensate, respectively. 
In principle, we could obtain the exact impurity operator by including an infinite number of possible excitations. However, in practice, this is only feasible for special cases such as the infinitely heavy impurity in a non-interacting quantum gas. 

\begin{figure}[t]
\begin{center}
\includegraphics[scale=0.5]{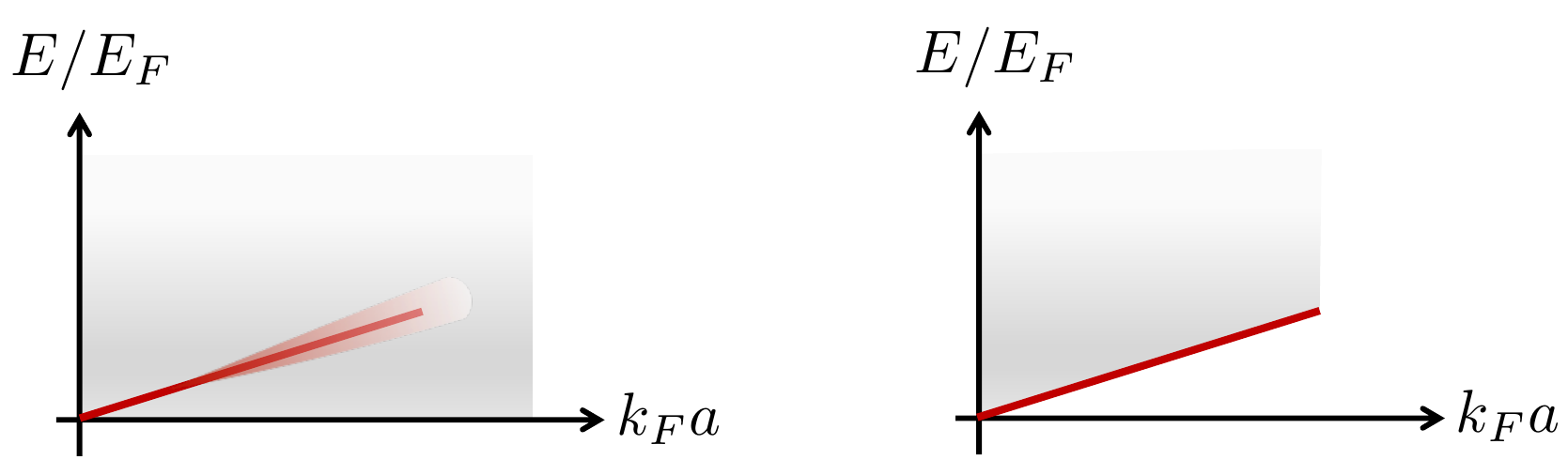}
\caption{Behavior of repulsive polarons for underlying attractive interactions (left) and purely repulsive interactions (right). In the former case, the polaron is embedded in a continuum (grey shading) at positive energies, thus resulting in broadening 
with increasing $k_F a$. By contrast, in the latter case, the continuum lies above the polaron energy and the polaron has an infinite lifetime.}
\label{fig:polaron-difference}
\end{center}
\end{figure}

To proceed, we consider the ``error operator'' $\hat\epsilon (t) = i \partial_t \hat{c}_0(t) - [\hat{c}_0(t),\hat{H}]$, which quantifies how well the approximate time-dependent impurity operator satisfies the Heisenberg equation of motion. We then construct the error quantity $\Delta (t) \equiv \tr\left[\hat\rho_{\rm med} \hat\epsilon (t) \hat\epsilon^\dag (t)\right]$, which we minimize with respect to $\partial_t\alpha^*_j(t)$ in order to obtain the differential equations for the parameters $\alpha_j(t)$.
Specializing to the Fermi polaron and taking the stationary condition $\alpha_j(t) = \alpha_j e^{-i\omega t}$, we finally obtain the coupled equations 
\begin{subequations} \label{eq:vareq}
\begin{align}
    \omega \alpha_0  & =  g n \alpha_0 + g\sum_{\k, \q}^\Lambda  (1-n_\k) n_\q \alpha_{\k\q}  \\ 
    \omega \alpha_{\k\q}  & =  \left[\epsilon_{\q-\k} + \ekmed - \epsilon_\q^{\rm med} + g n \right] \alpha_{\k\q} + g \alpha_0 \notag \\
    &+ g \sum_{\k'}^\Lambda (1-n_{\k'}) \alpha_{\k'\q}  - g\sum_{\q'}^\Lambda n_{\q'} \alpha_{\k\q'} ,
\end{align}
\end{subequations}
where we have used the simplified impurity-medium interaction defined by Eq.~\eqref{eq:1g1a}, and we have the Fermi-Dirac distribution for the medium particles
\begin{align}
    n_\k = \tr\left[\hat\rho_{\rm med} \hat{f}^\dag_\k \hat{f}_\k \right] = \frac{1}{1+e^{\beta (\ekmed-\mu)}} .
\end{align}

For the case of purely repulsive interactions, Eq.~\eqref{eq:1g1a} requires $g > 2\pi a/m_r >0$, and we can see in Eq.~\eqref{eq:vareq} that all the single-particle impurity energies are shifted upwards by $g n$, the Born approximation of the interaction energy.
Furthermore, $g n$ corresponds to an upper bound on the repulsive polaron energy in the ground state since $E \leq \bra{0} \hat{c}_0 \hat{H} \hat{c}^\dag_0 \ket{0} = gn$, where $\ket{0}$ is the state of the undisturbed  medium at zero temperature. Therefore, the lowest energy solution to Eq.~\eqref{eq:vareq} 
at zero temperature will always lie below the onset of the scattering continuum, and thus the variational approach yields a ground-state repulsive polaron with an infinite lifetime (see Fig.~\ref{fig:polaron-difference}). This energy gap between the polaron ground state and the onset of the continuum is an artifact of the approximation (similar to what was found for the attractive polaron~\cite{Trefzger2012}), since an exact theory with an infinite number of particle-hole excitations will contain excited polaronic states with energies arbitrarily close to the ground-state energy. 
Note that Eq.~\eqref{eq:vareq} correctly yields the mean-field energy $E \simeq 2\pi a n/m_r$ in the limit $n \to 0$, which one can show by making use of Eq.~\eqref{eq:1g1a}.

For attractive impurity-medium interactions, we can take the limit $g \to 0^-$ and still have a well-defined scattering length $a$, provided we also take $\Lambda \to \infty$. This corresponds to the case of a zero-range potential, which is a reasonable description for a dilute atomic gas where all the relevant length scales are much greater than the range of the van der Waals interactions. 
In this case, Eq.~\eqref{eq:vareq} is equivalent to the non-self-consistent $T$-matrix approach~\cite{Combescot2007,Liu2019}, where the impurity self energy $\Sigma=g\sum_{\k, \q}^\Lambda  (1-n_\k) n_\q \alpha_{\k\q}/\alpha_0$ corresponds to
\begin{align} \label{eq:selferg}
   & \Sigma(0,\omega) = \notag \\
   & \sum_\q n_\q \Bigl[\frac{m_r}{2\pi a}  - \sum_\k \Bigl( \frac{1-n_\k}{\omega-\epsilon_{\q-\k} - \ekmed + \epsilon_\q^{\rm med} + i0} \notag \\ 
   &\quad + \frac{1}{\ek + \ekmed} \Bigl) \Bigl]^{-1} . 
\end{align}
In contrast to purely repulsive interactions, we see here that there is no shift of the single-particle impurity energies and thus the self energy acquires an imaginary part when $\omega > 0$.
For Bose polarons above the BEC critical temperature, the self energy has the same form as Eq.~\eqref{eq:selferg}, where the Fermi distributions are instead replaced by Bose distributions, and we have $1+n_\k$ rather than $1-n_\k$ in Eq.~\eqref{eq:selferg}. 

\begin{figure*}[t]
\begin{center}
\includegraphics[scale=0.96]{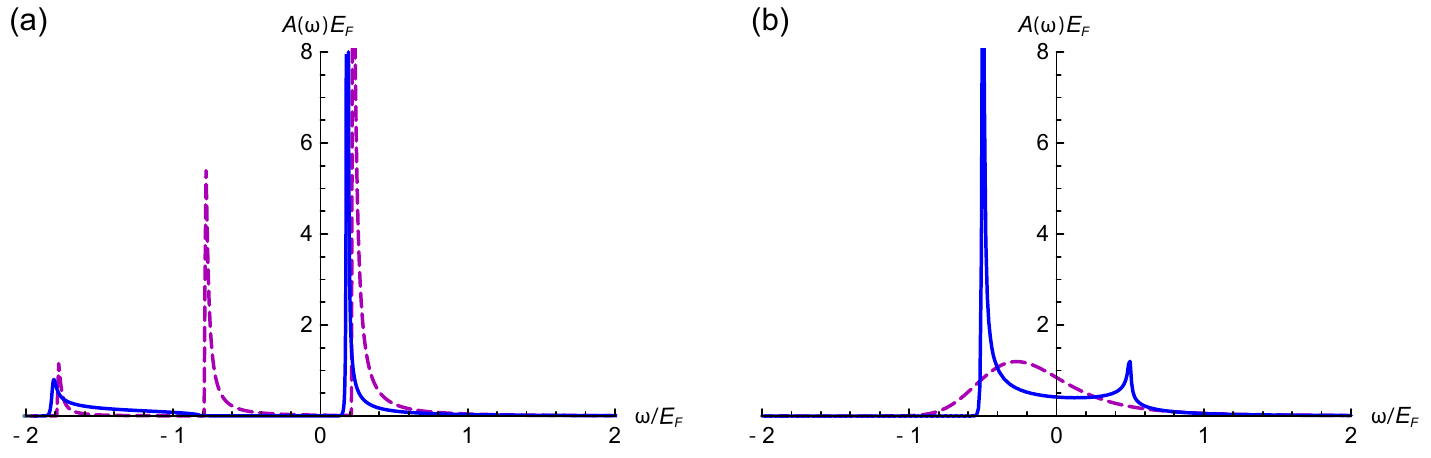}
\caption{Spectral function of an infinitely heavy impurity in an ideal Fermi gas at $T=0.01\,E_F$ (blue, solid) \cite{Goold2011} or an ideal Bose gas at $T=0$ \cite{Drescher2021} (purple, dashed). We show the result for $1/k_Fa=1$ (left) and $1/k_Fa=0$ (right). On the left, the two attractive peaks in the Bose case are due to single or double occupation of the impurity-boson bound state.}
\label{fig:spectrainf}
\end{center}
\vspace*{-10pt}
\end{figure*}

We can determine the lifetime of the metastable repulsive Fermi polaron in the weak-coupling limit by expanding Eq.~\eqref{eq:selferg} in $a$ and computing the lowest order contribution to the imaginary part~\cite{Adlong2020}. This yields, for arbitrary temperature, 
\begin{align}
    &\Gamma \simeq -  \Im[\Sigma(0,E)] 
    = \pi \left(\frac{2\pi a}{m_r} \right)^2 \times \\ 
    &\times \sum_{\k,\q} n_\q (1-n_\k) \delta\left(E-\epsilon_{\q-\k} - \ekmed + \epsilon_\q^{\rm med}\right) \hspace{5mm} {\rm (Fermi)}. \notag
\end{align}
Here, we see how $\Gamma$ is derived from the many-body continuum of states at positive energies, which is qualitatively distinct from the process of relaxation to lower energy states. 
Taking $T=0$, $m = m_{\rm med}$ and using the lowest order energy $E=4\pi a n/m$ then gives the 
expression for $\Gamma$ in Table~\ref{tab:pt}.
We stress that this imaginary contribution arising from many-body dephasing is not canceled by third-order diagrams involving self-energy insertions~\cite{Christensen2015}, since these only yield real terms.

The quasiparticle damping rate of the repulsive Bose polaron can be obtained in a similar manner, yielding the zero-temperature expression 
\begin{align}
    \Gamma \simeq 
    \pi n \left(\frac{2\pi a}{m_r} \right)^2 \sum_{\k} \frac{\ekmed}{E_\k} \delta\left(E-\epsilon_{\k} - E_\k \right) \hspace{5mm} {\rm (Bose)},
\end{align}
where the Bogoliubov dispersion $E_\k = \sqrt{\ekmed(\ekmed+2\mu)}$, and we have used the second-order self energy in Ref.~\citenum{Christensen2015}. In this case, $\Gamma$ depends on both the boson-boson and impurity-boson interactions. However, if we consider the special case where $a_B = a$ and $m = m_{\rm med}$, we obtain the simple expression in Table~\ref{tab:pt}. 

Finally, let us contrast the quasiparticle decoherence process described in this section with the standard Fermi liquid picture. 
There, the dominant mechanism for quasiparticle damping is momentum relaxation, which at low temperatures yields a lifetime proportional to $1/T^2$ \cite{Yan2019a}, and at zero temperatures to $1/p^4$ \cite{PinesNozieres,Bruun2008} for attractive Fermi polarons. 
However, the repulsive polaron has a finite lifetime at zero momentum, even in the absence of any decay into lower-lying excitations (such as attractive polarons, dressed molecules and tightly-bound dimers). In this sense, the repulsive polaron defies the standard Fermi liquid picture, which predicts that the lifetime should diverge at the impurity ``Fermi surface'' ($k=0$) \cite{PinesNozieres}.

\subsection{The case of an infinitely heavy impurity}

A useful reference system is that of an infinitely heavy impurity, which could in principle be approximately realized in cold-atom experiments by pinning the impurity atoms in tight traps. Remarkably, this problem affords analytic expressions for the spectral function when the medium is an ideal gas. Most notably, the Fermi polaron in the limit $m/m_{\rm med}\to\infty$ experiences the celebrated Anderson's orthogonality catastrophe~\cite{AndersonOC}, where the impurity ground states with and without interactions with the surrounding fermionic bath are orthogonal, i.e., the quasiparticle residue $Z=0$, and the spectrum features a power-law singularity, as quantified by the threshold behavior
\begin{align}
    A(\omega)\sim \theta(\omega-\omega_0)(\omega-\omega_0)^{\alpha-1}.
\end{align}
Here, the power law coefficient $\alpha=\delta(\kf)^2/\pi^2$ is related to the impurity-fermion scattering phase shift $\delta(k)=-\tan^{-1}(ka)$ evaluated at the Fermi surface, and the repulsive branch of the spectrum starts at 
$\omega_0=-\int_0^{\ef} dE\,\delta(\sqrt{2mE})/\pi$. 

More recently, the case of an infinitely heavy impurity in an ideal Bose gas has also been investigated~\cite{Drescher2021}. This is, of course, a highly singular limit. Even for a finite impurity mass, the lack of compressibility of the ideal Bose gas implies that $Z=0$ for the ground state~\cite{Shchadilova2016,Yoshida2018,Mistakidis2019,Guenther2021} and that the ground-state energy itself becomes ill defined for $a\geq0$. This highlights the crucial role played by boson-boson interactions in stabilizing the system~\cite{Levinsen2021}. In spite of this, the ideal Bose polaron spectrum remains well-defined, and in the case of an infinitely heavy impurity it can be obtained from an analytic solution of the impurity time evolution~\cite{Drescher2021}. The analysis of Ref.~\cite{Drescher2021} demonstrated a threshold behavior of the repulsive branch, valid when $0<k_Fa\ll1$,
\begin{align}
    A(\omega)\sim \theta(\omega-\omega_0)(\omega-\omega_0)^{-3/2}\exp\left(-\frac{8\pi n^2 a^4}{m(\omega-\omega_0)}\right),
\end{align}
where $\omega_0=2\pi an/m$. As in the Fermi case, the spectral function contains a power law, but in this case there is a strong suppression of spectral weight associated with low-energy excitations, preventing the power-law singularity.

The behavior in the two cases is illustrated in Fig.~\ref{fig:spectrainf}. One major difference is that the Fermi case features a spectral gap between the attractive and repulsive branches when $a>0$, whereas the Bose case instead has a strongly suppressed weight between the branches. Away from the resonance, the repulsive branches look reasonably similar; however as we approach the resonance, the behavior is remarkably different. We emphasize that in both the Fermi and Bose cases the problem is integrable, and completely determined by the solution of the two-body problem involving the impurity and a particle from the medium. Therefore, there is no analog of three-body recombination, and hence the population decay rate $\Gamma_p$ is identically 0. Likewise, our argument for the many-body dephasing relied on the existence of a well-defined quasiparticle that is pushed up into a scattering continuum, and thus also does not apply in either case. These points illustrate the strong qualitative differences between the mobile and the fixed impurity for properties beyond the impurity energy.

\section{Beyond the impurity limit: induced interactions and instabilities}
\label{sec:interactions}
\subsection{Polaron-polaron induced interactions}
When considering more than one impurity, the problem becomes significantly more complex at low temperatures where there are correlations between impurities. On the one hand, the statistics of the impurities starts to play a role. On the other, a variety of thermodynamic phases exist for the various components, which may be normal or condensed, mixed or phase-separated, {\it etc.}
Below we will treat separately the different cases which arise depending on the statistics of the bath and the impurities.

\subsubsection{Bosonic impurities in a Fermi sea} 
\label{subsubsec:boseImpsInFermiBath}
The energy density of a gas containing $N_\down$ bosonic impurities immersed in a bath of $N_\up\gg N_\down$ ideal fermions may be written as \cite{Viverit2000}
\beq
\mathcal{E}(n_\up,n_\down)=\frac{3}{5}E_F n_\up+E_\down n_\down+\frac{1}{2}F n_\down^2.
\eeq 
The various terms represent, in order, the (purely kinetic) energy of the unperturbed Fermi sea, the energy of isolated polarons, and the contribution due to \emph{polaron-polaron} interactions. We have omitted the mean kinetic energy of the impurities, which is negligible when cold bosonic impurities are considered.

The effective interaction $F$ between Landau quasiparticles can be split into two contributions: $F=g_1+F_{\rm x}$.  
The first is the direct interaction,  $g_1=4\pi a_{11}/m_{\downarrow}$, where $a_{11}$ is the scattering length between bare impurities (and we assume $|k_F a_{11}| \lesssim1$). 
The second term instead describes an exchange contribution, mediated by particle-hole excitations of the Fermi sea.

To obtain an explicit expression for the exchange interaction term, we follow the simple derivation outlined in Refs.~\cite{Yu2010,Yu2012}.  
An $\up$ atom and a $\down$ {\it polaron} interact with a coupling constant $g_{\rm x}$ given by 
\beq
g_{\rm x}=\frac{\partial^2\mathcal{E}}{\partial n_\up\partial n_\down}
=\frac{\partial \mu_\up}{\partial n_\down}
=\frac{\partial \mu_\down}{\partial n_\up}.
\eeq
Here and in the following, derivatives with respect to the density of a component will be taken at fixed density of the other.
To second order in $g_{\rm x}$, the polaron-polaron interaction is then given by 
\beq\label{E2_boseImps}
\mathcal{E}^{(2)}
=-\frac{g_{\rm x}^2}{V^3}\sum_{\k,\p,\q}
\frac
{(1-n^\up_{\k+\q})(1+n^\down_{\p-\q})n^\down_\p n^\up_\k}
{\ekpqmed + \ekpmq^* - \ep^*-\ekmed},
\eeq
where $n^\up_\k$ and $n^\down_\p$ indicate, respectively, Fermi and Bose distribution functions, since we are assuming bosonic impurities in a Fermi bath, and we have introduced the polaron dispersion $\ep^*=p^2/2m^*$.
The exchange contribution to Landau's polaron-polaron interaction is then obtained by differentiating with respect to the distribution functions of the two quasiparticles,
\beq\label{fx_boseImps}
F_{\rm x}=\frac{\delta^2\mathcal{E}^{(2)}}{\delta n^\down_{\p-\q}\delta n^\down_\p},
\eeq
where both $\p$ and $\q$ are assumed to be vanishingly small.
Performing the functional derivatives one finds 
\beq
F_{\rm x}=-\frac{g_{\rm x}^2}{V}\left(\sum_{\k}\frac{n^\up_\k-n^\up_{\k+\q}}{\ekpqmed -\ekmed}\right)_{q\rightarrow 0}=-g_{\rm x}^2L,
\eeq
where $L$ is the so-called Lindhard function \footnote{We follow here the sign convention used in Ref.~\cite{PethickSmithBook}, but  note that other sources define the Lindhard function with the opposite sign.}, which at zero temperature coincides with the density of states at the Fermi surface $\mathcal{N}=\left(\frac{\partial n_\up}{\partial \mu_\up}\right)=\frac{3n_\up}{2E_F}$. Collecting the above results, at zero temperature we obtain
\begin{align}
F_{\rm x}&=-g_{\rm x}^2\mathcal{N}
=-\left(\frac{\partial \mu_\up}{\partial n_\down}\right)^2 \frac{\partial n_\up}{\partial \mu_\up}
= -\left[-  \frac{\left(\frac{\partial \mu_\up}{\partial n_\down}\right)}{\left(\frac{\partial \mu_\up}{\partial n_\up}\right)}\right]^2\frac{\partial \mu_\up}{\partial n_\up}.
\end{align}
To simplify this expression, we use the {\it triple product rule} $\left(\frac{\partial x}{\partial y}\right)_z \left(\frac{\partial y}{\partial z}\right)_x \left(\frac{\partial z}{\partial x}\right)_y=-1$ to find 
\beq
-\left(\frac{\partial \mu_\up}{\partial n_\down}\right)_{n_\up}/\left(\frac{\partial \mu_\up}{\partial n_\up}\right)_{n_\down}
=
\left.\frac{\partial n_\uparrow}{\partial n_\downarrow}\right|_{\mu_\uparrow}
=
\Delta N,
\eeq
where $\Delta N$ is the number of bath particles in the dressing cloud of an impurity introduced in Eq.~\eqref{DeltaN}.
The latter expression shows that the induced quasiparticle interaction for bosonic impurities may be compactly written as 
\beq\label{fx_final}
F_{\rm x}= -\frac{\left(\Delta N\right)^2}{ \mathcal{N}}.
\eeq

This final result highlights the power and beauty of {\it Fermi liquid theory}: to derive the effective interaction we have used perturbation theory to describe the {\it weak} interaction mediated by the bath between two {\it quasi}-particles. The strong-coupling effects which generate the quasiparticles themselves are fully taken into account 
in Eq. (\ref{fx_final}) by means of $\Delta N$, a quantity which must be computed using a suitable theory describing the strongly coupled $N+1$ problem (like those described in the earlier sections of this work). As such, Eq.~\eqref{fx_final} holds for arbitrarily strong impurity-bath interaction strength provided there are no instabilities or transitions that invalidate the use of Fermi liquid theory.

Very recent measurements targeting locally-large concentrations of bosonic impurities in a Fermi bath showed a trend compatible with this Fermi liquid prediction \cite{DeSalvo2019,Fritsche2021}, but more accurate observations are needed to clearly pinpoint the phenomenon. Indeed, multiple issues render the measurements complicated in the case of a Bose-Fermi mixture. First, the relevant Feshbach resonances featured a relatively small magnetic field width, so that magnetic field instabilities generated large error bars. Second, the bosonic component is highly compressible, so that the density distributions can vary rapidly before and after the ``injection''. Indeed, for repulsive Bose-Fermi interactions, the mixture is highly unstable towards phase-separation \cite{Viverit2000,Lous2018,Huang2019}. On the other hand, the absence of Pauli blocking in the minority component means that it is possible to create mixtures featuring large local concentrations of impurities. As such, Bose-Fermi mixtures represent a very favorable setting for studying polaron-polaron induced interactions.

\subsubsection{Fermionic impurities in a Fermi sea} 
\label{subsubsec:fermiImpsInFermiBath}

\renewcommand{\labelenumi}{\roman{enumi}.}

When a fermionic bath hosts distinguishable fermionic impurities, the energy density may be written as
\beq\label{E_fermiImpsInFermiBath}
\mathcal{E}(n_\up,n_\down)=\frac{3}{5}E_F n_\up
+\mathcal{E}_{{\rm kin}\down}
+E_\down n_\down+\frac{1}{2}Fn_\down^2.
\eeq 
A number of differences are present with respect to the discussion of bosonic impurities given in the previous Sec.~\ref{subsubsec:boseImpsInFermiBath}:
\begin{enumerate}
    \item There is no direct interaction between identical fermionic impurities (i.e., $F=F_{\rm x}$).
    \item Pauli pressure dictates that impurities form their own Fermi sea, with Fermi energy $E_{F\down}=(n_\down/n_\up)^{2/3}(m/m^*)E_F$. The corresponding contribution $\mathcal{E}_{{\rm kin}\down}=\frac{3}{5}E_{F\down} n_\down$ to the energy density can be sizable, and this indeed permitted a direct measurement of the effective mass $m^*$ of the polarons via injection RF spectroscopy in Ref.~\cite{Scazza2017}.
    \item Correspondingly, final states available to the interacting impurities are ``Pauli blocked", rather than ``Bose enhanced", so that in the numerator of Eq.~\eqref{E2_boseImps} one needs to  replace $(1+n^\down_{\p-\q})$ by $(1-n^\down_{\p-\q})$, where $n^\down_{\k}$ is now a Fermi distribution function. As a consequence, the functional derivative with respect to the distribution functions of the minority particles in Eq.~\eqref{fx_boseImps} leads to an overall sign change in the exchange interaction term for fermionic impurities, which ultimately becomes repulsive and reads
\end{enumerate}
\beq
F_{\rm x}= \frac{\left(\Delta N\right)^2}{ \mathcal{N}}.
\eeq

Despite intense efforts, experiments on Fermi-Fermi mixtures have so far proved unable to unambiguously detect the presence of induced polaron-polaron interactions. Indeed, reported measurements remain compatible within the experimental uncertainties with a description in terms of uncorrelated quasiparticles up to sizable concentrations of impurities \cite{Schirotzek2009,Scazza2017}. Somewhat unexpectedly, the contribution $F n_\down^2/2$ due to induced interactions between quasiparticles  disappears from the equation of state when one switches from the canonical description used above to a grand-canonical formulation \cite{Mora2010}. In the latter case, the equation of state for the mixture gives a pressure which is the sum of the pressures of an ideal gas of fermions and an ideal gas of polarons, with no interaction terms. The apparent contradiction is resolved by realizing that the polaron energy is actually a function of the majority chemical potential $\mu_\up$, so that the two pressures are effectively coupled. A careful calculation shows that the correct interaction term $\propto F n_\down^2$ is recovered when switching back to the canonical ensemble.
A grand-canonical description is the appropriate one when extracting thermodynamic quantities from {\it in-situ} measurements on trapped gases, as done for example in Ref.~\cite{Nascimbene2010} following the proposal of Ref.~\cite{Ho2010}, but a canonical description is generally needed to describe RF experiments or QMC calculations, where a controlled number of impurities is present in the system.
The induced interaction term between polarons was indeed retained and shown to be important when analyzing state-of-the-art QMC calculations, and a complete phase diagram featuring the various possible phases arising in zero temperature Fermi-Fermi mixtures was derived in Ref.~\cite{Pilati2008}.

\subsubsection{Bosonic media} 
A bosonic bath is highly compressible, due to the absence of the large Fermi pressure. As a consequence, all bath particles condense in the same physical state at low temperatures, and the kinetic energy is negligible. On the other hand, the bath bosons experience a mutual direct (mean field) interaction. The energy density of the mixture may therefore be written as
\beq\label{E_fermiImpsInBoseBath}
\mathcal{E}(n_\up,n_\down)=\frac{2\pi a_B}{m}n_\up^2+\mathcal{E}_{{\rm kin}\down}+E_\down n_\down+\frac{1}{2}F n_\down^2,
\eeq
irrespective of the impurity statistics, where $\mathcal{E}_{{\rm kin}\down}=\frac35 E_{F\down}n_\down$ in the case of fermionic impurities, while it vanishes for bosonic impurities.
Even though the fundamental excitations in the bath are Bogoliubov modes, rather than particle-hole excitations, a calculation similar to that developed above leads to an identical result \cite{Yu2012,CamachoGuardian2018},
\beq
F_{\rm x}=\mp g_{\rm x}^2\frac{\partial n_\uparrow}{\partial \mu_\up} = \mp \frac{(\Delta N)^2}{\mathcal{N}}
\eeq
(with $-$ for bosonic impurities and $+$ for fermionic ones),  the only difference being that here $\mathcal{N}=\partial n_\up / \partial \mu_\up = m_{\rm med}/(4\pi a_B)$.

\begin{figure*}[t]
\begin{center}
\includegraphics[scale=1.33]{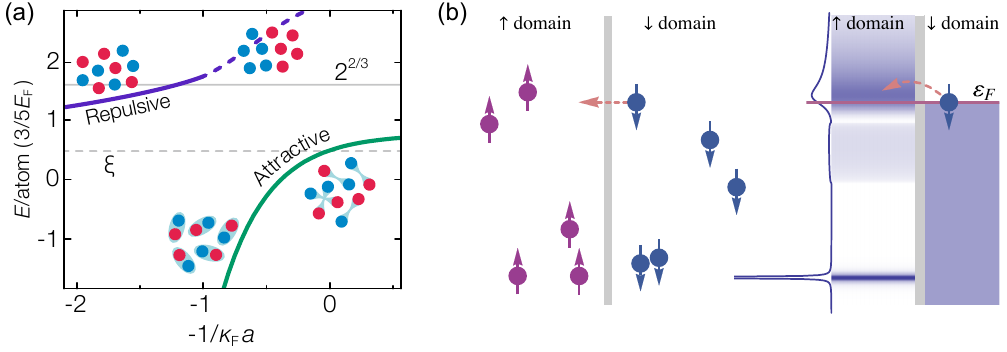}
\caption{Ferromagnetic instability of a repulsive balanced Fermi mixture. (a) The energy of the upper (purple) and lower (green) many-body branches is illustrated, corresponding to a net repulsive and attractive interaction, respectively. The ground state is a paired phase, featuring a mean energy per particle $E=\frac{3}{5}\xi E_F$ at $1/k_Fa=0$, where $\xi \approx 0.37$ is the Bertsch parameter \cite{Zwerger2011}. The repulsive Fermi liquid may undergo a ferromagnetic phase transition 
beyond a critical interaction strength along the upper branch, where a spin-segregated state is energetically favored. 
For an equal-mass, balanced mixture at $T=0$, this occurs when the mean energy per particle of the mixed, paramagnetic state exceeds that in a non-interacting Fermi gas, namely $E=\frac{3}{5}E_F$, by a factor $2^{2/3}$.
(b) Illustration of the stability condition for spin-polarized domains. A spin-$\downarrow$ fermion can tunnel across a spin domain wall, thereby becoming 
a repulsive quasiparticle within the spin-$\uparrow$ domain.
On the right, the energy levels available to the fermion are represented. If a well defined repulsive polaron exists and its energy overcomes the Fermi energy, i.e., $E>E_F$, tunneling through the interface is suppressed and the domain wall is energetically stable. Panel (b) is adapted with permission from Ref.~\citenum{NgampThesis}.}
\label{fig:FM}
\end{center}
\end{figure*}

\subsection{Ferromagnetic and pairing instabilities in Fermi-Fermi mixtures}
As mentioned in the introduction, a major motivation for studying the impurity problem is the fact that understanding quasiparticle properties of the polarized system provides important insight into the more challenging phase diagram of its population-balanced counterpart \cite{Parish2007}.
This holds true especially for the repulsive Fermi polaron, that was first considered \cite{Pilati2010,Chang2011} in connection with the physics of itinerant ferromagnetism of a repulsive Fermi gas, originally introduced by E. Stoner in his textbook model \cite{Stoner1933}. Repulsive polarons indeed constitute the building blocks of a repulsive Fermi liquid, which in turn represents the paramagnetic state of a two-component Fermi mixture with short-ranged inter-species repulsion. 

For a hypothetical, genuine two-body repulsive potential, a knowledge of the energy $E$ of repulsive polarons as a function of the interaction strength suffices to determine the emergence of a ferromagnetic instability in the $(N+1)$-particle system \cite{Pilati2010,Chang2011,Massignan_Zaccanti_Bruun}. This can be understood if one considers a bath of spin-$\uparrow$ electrons, interacting with a spin-$\downarrow$ electron impurity via a screened, short-range Coulomb repulsion. As long as the polaron energy does not exceed the Fermi energy $E_F$ of the surrounding bath, the spin-down quasiparticle embedded in the fermionic medium is energetically favored, and the system is in the Fermi liquid, paramagnetic state. Instead, for $E>E_F$, the system becomes unstable towards a fully ferromagnetic phase, as it is now energetically convenient for the impurity electron to flip its spin, resulting in a polarized Fermi gas of $N+1$ spin-$\uparrow$ particles: in this latter case, the energy cost of adding one spin-$\uparrow$ electron to the medium is lower than the energy cost associated with a strong impurity-bath repulsion. 
In contrast with the case of electrons in solids, where only the total electron population is fixed, in ultracold gases the two pseudospin numbers, $N_{\uparrow}$ and $N_{\downarrow}$, are generally fixed separately. As a result, the total “magnetization” $N_{\uparrow}- N_{\downarrow}$ is constrained by the two initial spin populations at which the gas is prepared. Yet, the same energetic argument applies \cite{Massignan_Zaccanti_Bruun}, and for $E>E_F$ ferromagnetism appears in this case as an instability of the repulsive Fermi liquid towards the formation of spatially separated, polarized domains of spin-$\uparrow $ or spin-$\downarrow $ particles (see Fig.~\ref{fig:FM}).

Such a seemingly simple scenario becomes richer and more complex once one considers the short-ranged attractive potentials relevant for a realistic description of the interaction between two ultracold atoms. 
As discussed in the previous sections, the presence of a lower-lying energy branch, connected with the existence of a weakly-bound molecular level below the two-body scattering threshold \cite{Chin2010}, makes repulsive polarons acquire a quasiparticle decay rate even at $T=0$.
As a consequence, ferromagnetism inherently competes with the tendency of the system to relax into the many-body ground state. 
Around the interaction strength $k_Fa \sim$ 1 relevant for ferromagnetism to develop, the spin-imbalanced ground state corresponds to phase separation between a superfluid of dimers and a spin-polarized Fermi gas~\cite{Pilati2008,Parish2021}. Therefore, here the attractive branch can be safely regarded as a paired state where all impurity particles are bound to a partner of the host gas~\cite{Schirotzek2009}.

In this framework, the importance of a proper interpretation of the polaron spectral width $\Gamma$ thus becomes clear:
If $\Gamma$ was exclusively linked to the population decay rate of the repulsive Fermi liquid \cite{Massignan_Zaccanti_Bruun}, one would expect the repulsive branch to become ill-defined at strong repulsion, and the physics of Stoner’s model to be inaccessible with such systems, since ferromagnetic correlations would be completely forestalled by attractive ones (such as pairing) \cite{Pekker2011}. 
Instead, in agreement with the recent theoretical analysis \cite{Adlong2020} outlined in Section~\ref{sec:qp-lifetime}, experimental pursuits have provided convincing evidence for the metastability of the repulsive branch even at strong coupling \cite{Kohstall2012,Scazza2017,Oppong2019}. In particular, a short but finite time window was identified within which the instability of the repulsive Fermi gas towards a magnetically correlated state could be observed. Spectroscopic studies on $^6$Li mixtures in the impurity limit indeed revealed well-defined quasiparticles even at very large repulsion, where $E$ exceeds the Fermi energy of the bath while $m^*$ diverges, indicating both an energetic and thermodynamic instability of the repulsive Fermi liquid state beyond critical interactions \cite{Scazza2017}. 

Studies of spin-dynamics in a repulsive Fermi gas \cite{Valtolina2017} artificially initialized in two spin domains separated by a ferromagnetic domain wall  
further allowed the characterization of the ferromagnetic behavior of the system, including the experimental determination of the metastability region of the ferromagnetic state in the temperature-interaction plane. 
This was linked to the softening of the spin-dipole mode, in quantitative agreement with theory models \cite{Recati2011, Grochowski2017} that completely neglect pairing correlations. Finally, later works based on time-resolved pump-probe spectroscopic techniques 
succeeded in tracing the out-of-equilibrium dynamics of two-component balanced spin mixtures \cite{Amico2018}. For this, fermions were selectively brought to strong repulsion along the upper branch of a Feshbach resonance, quickly enough to avoid undergoing substantial dynamics during the preparation itself, starting essentially from the paramagnetic Fermi liquid state. The rapid growth of short-range anti-correlations between repulsive quasiparticles was observed beyond critical interactions, demonstrating that concurrent pairing processes could be initially overcome. These studies \cite{Amico2018}, paralleled also by spin-density noise correlation measurements and monitoring of other macroscopic observables \cite{Scazza2020}, led to the discovery of an unpredicted, emergent heterogeneous phase: a quantum emulsion where paired and unpaired fermions macroscopically coexist, while featuring microscale phase separation. In turn, this observation links the physics of the repulsive Fermi gas to certain strongly correlated electron materials, where competing order parameters coexist in nanoscale phase separation \cite{Dagotto2003, Dagotto2005}.

\section{Concluding remarks}

Repulsive mixtures of ultracold atoms, both within and beyond the impurity limit, feature a very rich and intriguing phenomenology. 
Firstly, owing to the metastability of the upper many-body branch, whose population decays on a timescale significantly slower than that set by quasiparticle decoherence, Fermi-Fermi mixtures are suited to investigate some aspects of Stoner’s model \cite{Stoner1933}, enabling one to gain important insights into the magnetic properties of itinerant fermionic particles \cite{Valtolina2017,Amico2018,Scazza2020}. 
The presence of a concurrent pairing instability in these systems, with which ferromagnetism inherently competes, further enriches this scenario, triggering non-trivial dynamics over longer timescales and promoting the emergence of unpredicted heterogeneous many-body states \cite{Amico2018,Scazza2020}. In future experiments, it would be interesting to explore the transport properties of such a spatially inhomogeneous, slowly relaxing state, and to probe its emergence in box-like potentials \cite{Navon2021}, weak optical lattices \cite{Pilati2014, Zintchenko2016} or lower dimensions \cite{Conduit2010,Cui2014}. Quantum gas microscopes \cite{Gross2017} could uniquely explore the competition between antiferromagnetic ordering, favored by the underlying lattice structure, and quantum emulsions of itinerant fermions. Further, the realization of such a rich scenario opens exciting possibilities to dynamically create elusive phases of magnetized superfluidity \cite{Casalbuoni2004, Fukushima2010} and to spontaneously attain mesoscopic magnetic impurities within strongly interacting superfluids \cite{Magierski2019}. 

Secondly, embedding highly-compressible bosonic impurities in a large Fermi sea \cite{Fritsche2021} paves the way for the observation of a crucially missing milestone in the experimental demonstration of the Fermi liquid paradigm in ultracold atoms, namely, polaron-polaron interactions. The latter are predicted to be even stronger between Bose polarons, since a dilute bosonic bath is highly compressible and can more easily convey excitations \cite{CamachoGuardian2018}. 

In the future, Bose and Fermi polaron studies with ultracold gases could be extended to the case where the motions of the impurity and/or the medium are confined within a lattice potential \cite{Knap2012}, possibly in reduced dimensions and in the presence of engineered site-resolved disorder \cite{Hu2016a}. For a comprehensive modeling of such scenarios, an essential preliminary step is to derive \emph{ab initio} the effective interaction parameters associated with two atoms interacting strongly in an optical lattice potential, extending previous three-dimensional calculations \cite{Buechler2010}. 
The probing techniques discussed in this review could also find potential applications in the investigation of magnetic polarons \cite{Ashida2018,Koepsell2019,Ji2021}, i.e., charge excitations moving within a magnetic background where they become dressed and acquire renormalized quasiparticle properties \cite{Alexandrov2010,Kane1989}. In particular, such excitations emerge in fermionic Hubbard models from the interplay between charge and spin \cite{Grusdt2018a}, as observed by experiments \cite{Koepsell2019,Ji2021}. These recent observations have attracted much interest \cite{Grusdt2019,Blomquist2020,Nielsen2021}, since the dynamics of magnetic polarons in doped Mott insulators may explain the behavior of certain strongly correlated materials such as high-temperature superconductors \cite{Lee2006}. 
Indeed, the strong analogies between the polaron quasiparticles realized in ultracold atomic mixtures and in the solid state could contribute to progress along various directions of research. In particular, there are now prospects to realize Bose polarons with exciton-polaritons in semiconducting microcavities~\cite{Levinsen2019}, which could allow one to enhance photon-photon correlations.


\begin{acknowledgements}

The authors would like to thank Luis Pe\~na Ardila for providing the data from Ref.~\citenum{Ardila2018}, Richard Schmidt for providing the data from Ref.~\citenum{Schmidt2011}, Moritz Drescher for providing data from Ref.~\citenum{Drescher2021}, and Haydn Adlong and Weizhe Liu for providing additional data for Fig.~\ref{fig:spectrainf}.
We also thank 
Haydn Adlong,
Georg Bruun,
Fr\'ed\'eric Chevy,
Tilman Enss, 
Weizhe Liu,
Brendan Mulkerin,
Alessio Recati,
Giacomo Roati, 
and Richard Schmidt for insightful discussions.

F.S. acknowledges funding from the European Research Council (ERC) under the European Union’s Horizon 2020 research and innovation programme (Grant Agreement no.~949438). M.Z. acknowledges funding from the ERC through grant no.~637738 PoLiChroM and from the Italian MIUR through the FARE grant no.~R168HMHFYM.
P.M. was supported by grant PID2020-113565GB-C21 funded by MCIN/AEI/10.13039/501100011033, by EU FEDER Quantumcat, by the National Science Foundation under Grant No. NSF PHY-1748958, and by the {\it ICREA Academia} program.
J.L. and M.M.P. acknowledge support from the Australian Research Council Centre of Excellence in Future Low-Energy Electronics Technologies (CE170100039). J.L. and M.M.P. are also supported through the Australian Research Council Future Fellowships FT160100244 and FT200100619, respectively, and J.L. furthermore acknowledges support from the Australian Research Council Discovery Project DP210101652.
\end{acknowledgements}

\bibliography{bib}

\end{document}